\UseRawInputEncoding
\documentclass[aip,amsmath,amssymb,reprint]{revtex4-1}

\usepackage{epsfig}
\usepackage{latexsym}
\usepackage{amsmath}
\usepackage{amsfonts}
\usepackage{amssymb}
\usepackage{amsbsy}
\usepackage{subfigure}
\usepackage{bm}
\usepackage{color}
\usepackage{dcolumn}
\usepackage{hyperref}

\usepackage{graphicx}

\DeclareMathOperator{\arccot}{arccot}

\graphicspath{{figures/}}

\newcommand{\lla}{\left\langle}
\newcommand{\rra}{\right\rangle}

\newcommand{\vareps}{\varepsilon}

\begin{document}
\title{Wall-anchored semiflexible polymer under large 
amplitude oscillatory shear flow}
\author{Antonio Lamura}
\email[]{Corresponding author: antonio.lamura@cnr.it}
\affiliation{
Istituto Applicazioni Calcolo, CNR,
Via Amendola 122/D, 70126 Bari, Italy}
\author{Roland G. Winkler}
\email[]{r.winkler@fz-juelich.de} \affiliation{Theoretical Physics of Living Matter, Institute for Advanced Simulation and Institute of Biological Information Processing,
Forschungszentrum J\"{u}lich, 52425 J\"{u}lich, Germany}
\author{Gerhard Gompper}
\email[]{g.gompper@fz-juelich.de} \affiliation{Theoretical Physics of Living Matter, Institute for Advanced Simulation and Institute of Biological Information Processing,
Forschungszentrum J\"{u}lich, 52425 J\"{u}lich, Germany}
\date{\today}
\begin{abstract}
The properties of semiflexible polymers tethered by one end to an impenetrable wall and exposed to oscillatory shear flow are investigated by mesoscale simulations. A polymer, confined in two dimensions, is described by a linear bead-spring chain, and fluid interactions are incorporated by the Brownian multiparticle collision dynamics approach. At small strains, the polymers follow the applied flow field. However, at high strain, we find a strongly nonlinear response, with major conformational changes. Polymers are stretched along the flow direction and exhibit U-shaped conformations while following the flow. As a consequence of confinement in the half-space, a frequency doubling  in the time-dependent polymer properties appears along the direction normal to the  wall.   
\end{abstract}
\maketitle

\section{Introduction}

Studies of the nonequilibrium behavior of polymer systems provide a link between their microscopic molecular characteristics and the emerging macroscopic dynamical properties \cite{bird:87.1,lars:99,rubi:03,lars:15,shaw:18,schr:18}. 
A major experimental breakthrough in resolving and visualizing the nonequilibrium properties of individual molecules was achieved by studies on single DNA filaments \cite{perk:95}, which  paved the way to explore a large variety of polymer properties  under flows. 

From the theoretical side, several detailed computational molecular models have been considered,
which enhance our understanding of molecular processes. In particular,  
single-polymer studies provide the opportunity to directly observe the microscopic
conformational properties  of individual polymers close to equilibrium or under flow conditions, thereby facilitating access to their non-equilibrium conformational properties, which are
ultimately responsible for the macroscopic rheological behavior.
The desire to visualize  individual polymer conformations in flow, both from an experimental and simulation point of view, is strongly linked with advances in the statistical description of their properties provided by molecular
theoretical models \cite{schr:18,wink:06.1,wink:10}. 

Semiflexible polymer-type molecular structures are omnipresent in biological systems, e.g., DNA, actin filaments, and  microtubules, and  their rigidity is fundamental for  their functions. Indeed, actin
filaments contribute with their rigidity to the mechanical properties of the cytoskeleton and the ability of DNA to pack in the genome or inside a virus capsid is controlled by its persistence length. As a consequence, the properties of semiflexible polymers have been intensively investigated
\cite{wilh:96,goet:96,harn:96,ever:99,samu:02,lego:02,wink:03,petr:06}.
Here,  theoretical \cite{bird:87,oett:96,wink:06_1,munk:06,wink:10} and computational
\cite{hur:00,jend:02,hsie:04,liu:04,cela:05,ryde:06,ripo:06,send:08,he:09,
zhan:09,huan:10,huan:11,huan:12,lang:14,niko:17,kong:19,romo:21,
shee:21,niko:21} studies revealed novel dynamical, conformational, and rheological properties.

The large majority of these studies focuses on three-dimensional systems, and predominantly single-polymer theoretical and simulation studies have been performed  under steady shear and extensional  flow  in the bulk \cite{shaq:05,wink:10,huan:11}. Similarly, in experiments single DNA filaments have been considered, in particular filaments grafted on a wall by one end under steady shear, and their mechanical properties  \cite{smit:92} and  relaxation dynamics \cite{perk:94} 
have been examined. Further experiments \cite{doyl:00,luet:09} revealed the existence of a cyclic dynamics, which has been confirmed by simulations \cite{luet:09,delg:06}. 

Polymers in two dimensions are of interest on their on. Compared to the three-dimensional case, there are
two major differences. On the one hand, excluded-volume interactions are more important and, on the other hand, hydrodynamics can be neglected, specifically,  in the case of strongly adsorbed
polymers \cite{maie:02}. Such systems can be realized experimentally  by considering filaments
strongly adsorbed onto a surface, a membrane, or at interfaces separating
immiscible fluids \cite{maie:02,cher:14,hero:10}.  In particular,  end-anchored semiflexible polymers have been considered,  where the central monomer is subject to an oscillatory force, and a transition from a limit cycle to an aperiodic dynamics with increasing rigidity has been predicted theoretically \cite{chat:07}. The dynamics of semiflexible polymers under 
shear has been  investigated numerically and substantial stiffness-dependent  conformational changes at high flow rates have been 
obtained \cite{lamu:12}. Specifically, more flexible polymers are extended by the flow, while stiffer filaments  contract. Moreover, filaments are aligned by the flow with a tumbling motion that, at high shear rates, resembles the motion of flexible
polymers in three dimensions.

Theory, simulations, and experiments typically consider stationary flows in the linear viscoelastic regime.  However, it is important to unravel the polymer dynamics when time-dependent flows are applied, specifically large amplitude oscillatory shear (LAOS) flows. 
Such studies provide  additional insight into the  macroscopic polymer properties, specifically their viscoelastic behavior \cite{hyun:11,roge:18}. 
Despite the deep interest and experimental relevance, so far  very few theoretical and simulation  studies have been performed under
such flow conditions. In Ref.~\cite{chen:05}, the extension and migration of
a chain, confined in a microchannel and subject to oscillatory pressure-driven flow, were observed. Later,  Brownian dynamics studies of macromolecules under oscillatory shear flow have been performed \cite{thom:09,lamu:19}.
In contrast, many more  LAOS experiments have been conducted \cite{hyun:11,zhou:16,roge:18}.
Yet, no study so far considered a single tethered semiflexible chain subject to oscillatory shear flow. 

In this paper, we investigate the conformational and dynamical properties  of a semiflexible polymer in two dimensions,  tethered by one of its ends on a reflecting wall, in the presence of an oscillatory shear flow by mesoscale simulations. A polymer is modeled as a self-avoiding  worm-like filament. Neglecting hydrodynamic interactions, the polymer  is assumed to be in contact with a Brownian heat bath implemented via the Brownian version \cite{ripo:07,gomp:09} of the multiparticle collision dynamics approach \cite{kapr:08,gomp:09}.
The polymer properties are characterized in terms of the strain defined as the ratio of the shear rate to
the shear frequency. For low strain, the polymers behave roughly as at equilibrium. In the opposite limit of high strain, they show in each shear half-cycle average properties comparable to those at steady shear---they  fully elongate and flip back and forth following the external flow. For intermediate values of strain,  flipping is disfavored, and  polymers may remain preferentially in a coiled state when shear changes sign without re-orientating along the instantaneous flow direction. The analysis of the amplitudes of the undulation modes suggests that polymers exhibit a flexible polymer-like behavior at larger length scales  along the shear
direction, despite their significant stiffness in absence of the external flow.  The periodic motion of the center-of-mass displays the same frequency as the external shear along the flow direction, while frequency doubling in the normal direction  appears by  wall reflection, which ``repels'' the polymer as flow drags it  from one side to the other in a cycle.

The numerical models for the polymer and the Brownian fluid are introduced in  Sec.~\ref{sec:model}. The results for the conformational and the dynamical behavior are presented in Sec.~\ref{sec:results}.
Finally, Sec.~\ref{sec:conclusions}  summarizes and discusses the main findings
of this study.

\section{Model and Method} \label{sec:model}

We model a filament as a linear bead-spring chain with $N$ beads of mass $M$ separated by bonds of length $r_0$, and  confined in the positive half-plane of the two dimensional space. The first bead is tethered at the origin $(0,0)^T$
of the Cartesian reference frame, with no preferred orientation of the first bond. The beads are subjected to forces by the total potential
$U=U_{bond}+U_{bend}+U_{ex}+U_w$.
Bonds are described by the harmonic potential
\begin{equation}\label{bond}
U_{bond}=\frac{\kappa_h}{2} \sum_{i=1}^{N-1}
(|{\bm r}_{i+1}-{\bm r}_{i}|-r_0)^2 ,
\end{equation}
where ${\bm r}_i$ denotes the position vector of the $i-$th bead
($i=1,\ldots,N$) and $\kappa_h$ is the force constant.
The stiffness of the polymer is implemented by the bending potential
\begin{equation}
U_{bend}=\kappa \sum_{i=1}^{N-2} (1-\cos \varphi_{i}) ,
\label{bend}
\end{equation}
where $\kappa$ is the bending rigidity and $\varphi_{i}$ is the angle between two consecutive bond vectors. The filament persistence
length $L_p$ is related to $\kappa$ via $L_p=2 \kappa r_0/ k_B T$, where
$k_B T$ is the thermal
energy, with $T$ the temperature and $k_B$ Boltzmann's constant.
The shifted and truncated Lennard-Jones potential
\begin{equation}
U_{ex} =
4 \epsilon \Big [ \Big(\frac{\sigma}{r}\Big)^{12}
-\Big(\frac{\sigma}{r}\Big)^{6} +\frac{1}{4}\Big] \Theta(2^{1/6}\sigma -r)
\label{rep_pot}
\end{equation}
ensures the self-avoidance of non-connected beads. Here,
$r$ is the distance between two beads and
$\Theta(r)$ is the Heaviside function ($\Theta(r)=0$ for $r<0$ and
$\Theta(r)=1$ for $r\ge 0$).
Confinement in the half-plane $y > 0$ is achieved by
a repulsive wall implemented via the potential
\begin{equation}
U_{w} =
4 \epsilon \Big [ \Big(\frac{\sigma_w}{y}\Big)^{12}
-\Big(\frac{\sigma_w}{y}\Big)^{6} +\frac{1}{4}\Big] \Theta(2^{1/6}\sigma_w -y) ,
\label{rep_pot2}
\end{equation}
where $y$ is the distance of a bead from the wall.
The dynamics of the beads is described by Newton's equations of motion, which are  integrated by
the velocity-Verlet algorithm with time step $\Delta t_p$
\cite{swop:82,alle:87}.

Shear flow and thermal fluctuations are
implemented by  the Brownian multiparticle collision dynamics  approach (B-MPC)
\cite{ripo:07,gomp:09,male:00_1}, where hydrodynamic interactions are  neglected.
In this method, stochastic collisions between each bead 
and a number $\rho$ of ``fluid'' phantom  particles of mass $m$
mimic interactions of a
fluid volume surrounding a bead. The moment of a phantom
particle is Maxwellian  distributed, with variance $m k_B T$ and mean  $(m\dot{\gamma}
y \sin(2 \pi t/T),0)^T$ at time $t$ in the presence of the oscillating shear flow
of shear rate $\dot \gamma$, period $T$, and orientation along the $x$-axis.
The collision process itself is implemented via the stochastic rotation dynamics (SRD) variant
of the MPC method \cite{ihle:01,lamu:01,gomp:09}. Here, the
relative velocity of a bead, with respect to the
center-of-mass velocity of the bead and its related phantom
particles, is rotated in the $xy-$plane by a fixed angle  $\pm \alpha$ of uniformly distributed sign.
Collisions occur at time intervals $\Delta t$, where  $\Delta t > \Delta t_p$.

Simulations  are performed for the  parameters: $\alpha=130^{o}$,
$\Delta t=0.1 t_u$, with the time unit $t_u=\sqrt{m r_0^2/(k_B
T)}$, $M =\rho m$, $\rho=5$, $\kappa_h r_0^2/(k_B T)=4 \times 10^3$,
$\epsilon /
(k_B T)=1$, $\sigma=r_0$, $\sigma_w=r_0/2$, $N=101$,
and $\Delta t_p=10^{-2} \Delta t$. The value of
$\kappa_h$ ensures the length $L=100r_0$ of the polymer  within
$1\%$ for all systems and flow conditions.

\section{Results} \label{sec:results}

We study semiflexible polymers with the persistence lengths
$L_p/L=0.5$ and $2$. B-MPC simulations of free filaments yield the  end-to-end vector relaxation times
$\tau_r\simeq (1.9, 3.9) \times 10^6 t_u$ 
\cite{lamu:12}.
The tethered polymers are initialized with beads aligned along the
$y$-direction, and are equilibrated up to  $5 \times 10^6 t_u$. Subsequently,
data are collected up to the longest simulated time $10^{8} t_u$ and
averaged over half-periods with positive shear flow.
The flow is characterized in terms of the Weissenberg number
$Wi=\tau_r \dot{\gamma}$ and the Deborah number
$De=\tau_r \omega$, where $\omega=2 \pi /T$ is the shear frequency.
Since in oscillatory flow the strain
in a half-cycle is proportional to $\dot{\gamma}/\omega=Wi/De$,
averages will be characterized as function
of this ratio in the following.
The shear rate is varied the range 
$0 \leq Wi \lesssim 10^3$,
while the frequency is altered such that $0.1 \leq Wi/De \leq 100$.

\subsection{Conformational properties} \label{subsec:conform}

We characterize the filament conformational properties by the end-to-end vector ${\bm R}_e={\bm r}_N-{\bm r}_1$. The oscillatory shear implies a cyclic dynamics of filament collapse, stretching, and alignment along the flow direction (see
movies movie1.mp4, movie2.mp4, movie3.mp4, and movie4.mp4 in supplementary
material.).

\subsubsection{Filament stretching}

Stretching along the flow direction is characterized by the maximum extension $R_{ex,max}$ of the polymer
along the $x$-axis in a cycle, or, equivalently, by the average deficit length-ratio $\vareps =1-\lla R_{ex,max} \rra/L$, where the average is performed over periods. 

\begin{figure}[ht]
\begin{center}
\includegraphics*[width=\columnwidth,angle=0]{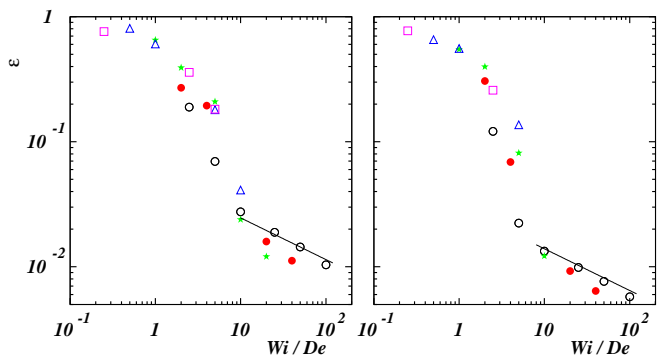}
\caption{Mean deficit length-ratio $\vareps$ 
as a function of the strain $Wi/De$
for $L_p/L=0.5$ (left), $2$ (right) and
$De=10$ ($\circ$, black), $25$ ($\bullet$, red), $50$ ($\star$, green), $100$ ($\triangle$, blue), and $200$ ($\square$, magenta). The slope of the full lines is $-1/3$.
\label{fig:deficit}
}
\end{center}
\end{figure}

Figure~\ref{fig:deficit} displays $\varepsilon$
as a function of the strain $Wi/De$. The data for the various $De$ nearly collapse when plotted as function of $Wi/De$, a feature applying to other quantities too.  Three regimes can be identified, which we will denote as low- ($Wi/De \lesssim 1$), intermediate- ($1 \lesssim Wi/De \lesssim 10$), and high-strain regime ($Wi/De \gtrsim 10$).
In the low-strain regime, either the shear rate or/and the period are so small that the shear is hardly able to deform the polymer compared to the equilibrium value.  With increasing strain beyond unity,
the deficit length decreases very rapidly and for $Wi/De >10$ a power-law regime seems to appears,   
with an exponent of $-1/3$ for $De=10$. This exponent is consistent with experimental results  
on tethered semiflexible DNA molecules under steady shear \cite{lado:00}. 
The high-strain regime corresponds to high values either of the shear rate
or the shear period, implying a large polymer stretching ($\vareps \lesssim 10^{-2}$), and the approximations of Ref.~\cite{lado:00} in the
derivation of shear-rate dependence apply. 
For  $De >10$, there seem to be some deviations from the
power law $-1/3$, or the power-law regime has not yet been reached;  
a detailed analysis is hampered by necessary  very large shear rates.
Noteworthy, stiffer polymers are stronger stretched than flexible ones. 

\subsubsection{Height above wall}

\begin{figure}[ht]
\begin{center}
\includegraphics*[width=\columnwidth,angle=0]{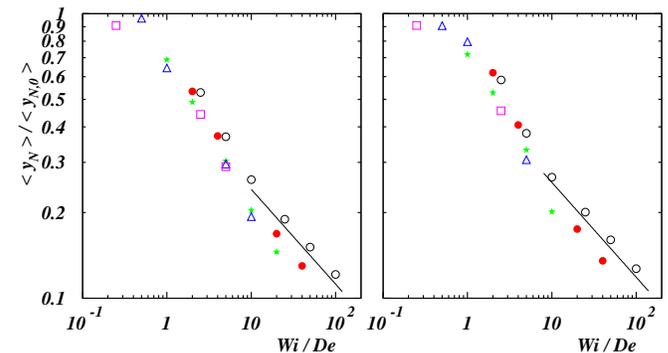}
\caption{Mean polymer height above wall as a function of the strain $Wi/De$
for $L_p/L=0.5$ (left), $2$ (right), and
$De=10$ ($\circ$, black), $25$ ($\bullet$, red), $50$ ($\star$, green), $100$ ($\triangle$, blue), and $200$ ($\square$, magenta).  The solid lines indicate the slope $-1/3$.
$\lla y_{N,0} \rra$ is the mean polymer height at equilibrium.
\label{fig:height}
}
\end{center}
\end{figure}

Confinement to the positive semi-space breaks spatial symmetry and leads to a geometry-induced polymer stretching normal to the wall.
The dependence of the height above the wall, defined as the average distance $\lla y_{N} \rra$ of bead $N$ from
the wall, on strain is depicted in Fig.~\ref{fig:height}. 

The height decreases with increasing strain according to the power-law  
$\lla y_{N} \rra \sim (Wi/De)^{-1/3}$ for high strain, independent of the considered stiffness. The range of the scaling regime roughly coincides with that in Fig.~\ref{fig:deficit} for the deficit length. Moreover, our simulations yield 
the dependence $\lla y_{N}^2 \rra \sim (Wi/De)^{-2/3} \sim 
\lla y_{N}\rra^2$ 
in the high $Wi/De$ regime. 
This dependence agrees with the scaling prediction for the fluctuations normal to a wall 
of a free-draining semiflexible polymers under shear flow \cite{lado:00}.

\subsubsection{Polymer deformation}

\begin{figure}[ht]
\begin{center}
\includegraphics*[width=\columnwidth,angle=0]{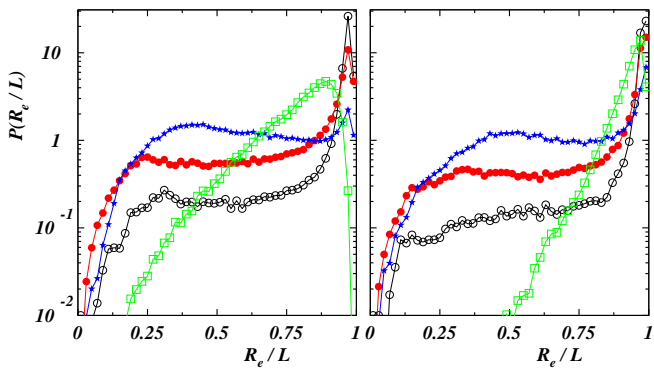}
\caption{Probability distribution function of the polymer end-to-end distance
$|\bm R_e|$ for $L_p/L=0.5$ (left) and $2$ (right), and the Weissenberg number $Wi=0$ ($\Box$, green) as well as $Wi=500$ for
$De=10$ ($\circ$, black), $50$ ($\bullet$, red), $200$ ($\star$, blue).
\label{fig:pdflen}
}
\end{center}
\end{figure}

More detailed insight onto the polymer deformation is gained by the distribution function of the 
end-to-end vector $R_e = |\bm R_e|$ (see Fig.~\ref{fig:pdflen}). 
In the absence of shear,  the distribution function exhibits a single peak at $R_e/L \simeq 1$ for $L_p/L=2$
corresponding to a fully elongated configuration. With decreasing  persistence length,
the distribution broadens and the peak moves to $R_e/L \simeq 0.87$. 
For the stiffer chain at small strain, $Wi/De=2.5$, the conformations are already substantially affected by shear, and $R_e$ values over a broad range, $0.3 \lesssim |\bm R_e/L| \lesssim 0.85$, are nearly equally probable. This range broadens with decreasing $De$ at fixed $Wi$, the probabilities in the plateau-like regime decrease, and a peak appears in the vicinity of full stretching. The latter indicates an increasing  probability of strongly-stretched polymers. At the same time, $P(R_e)$ decreases at smaller end-to-end vectors. We obtain approximately the same distribution functions for the more flexible polymer. 

The variations of the distribution function with $De$ affect the average end-to-end vector $R_e$.
Associated with the overall decrease of the distribution function by a change of $Wi/De$ from zero to $Wi/De=2.5$, is a small drop of the average mean end-to-end distance, $\lla R_e \rra$. A further increase of $Wi/De$ implies a monotonic swelling of  $\lla R_e \rra$. 
At the same time, $\lla (R_e - \langle R_e\rangle)^2 \rra$ swells first with increasing strain, reaches a maximum at $Wi/De \approx 8$, and decreases then by a  power-law with the exponent $-1/3$.

\subsection{Dynamical properties} \label{subsec:dynamics}

\begin{figure}[ht]
\begin{center}
\includegraphics*[width=0.8\columnwidth,angle=0]{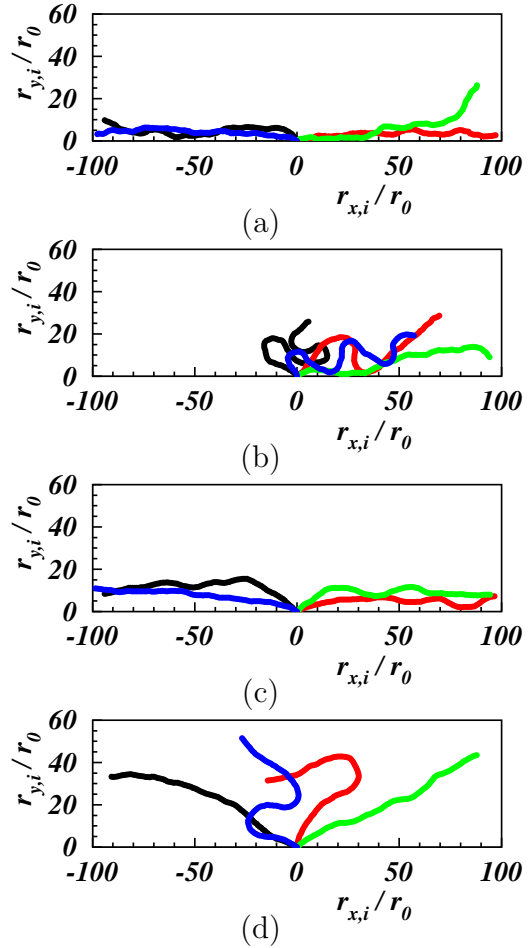}
\caption{Polymer conformations at the times $t/T= 1, 1.25, 1.5, 1.75$ 
(black, red, green, and blue lines, respectively) 
after equilibration
for $L_p/L=0.5$ ((a), (b)) and $L_p/L=2$ ((c), (d)),   $Wi=500$, and 
$De=10$ ((a), (c)) and $200$ 
((b), (d)). 
(See movies movie1.mp4 for the case (a), 
movie2.mp4 for the case (c),
movie3.mp4 for the case (b), 
and movie4.mp4 for the case (d) in supplementary material.)
\label{fig:conf05}
}
\end{center}
\end{figure}

Figure~\ref{fig:conf05} displays conformations of polymers during one period.
At high strain, a polymer flips back and forth following the applied shear. However,
the flipping mechanism strongly depends on the interplay of shear and
bending rigidity. In the case of a more flexible polymer, the polymer recoils after
the flow changes sign, then flips and elongates (see supplementary movie 
movie1.mp4).
In contrast, the stiffer polymer bends, assumes a U-shaped conformation during flipping, and finally elongates
(see the supplementary movie movie2.mp4). 
Smaller strain disfavors chain flip when the flow changes sign. Here, 
a polymer remains coiled without flipping (see Fig.~\ref{fig:conf05} (b) and the supplementary movie movie3.mp4), and, consequently,  remains essentially in the positive half-space $x>0$. 
This behavior is also observed for higher bending stiffness. In the latter case, however, the
chain flip occurs more frequently than in the flexible case with the
polymer recoiling and reorienting (see Fig.~\ref{fig:conf05} (d)
and the supplementary movie movie4.mp4).

\begin{figure}[ht]
\begin{center}
\includegraphics*[width=0.7\columnwidth,angle=0]{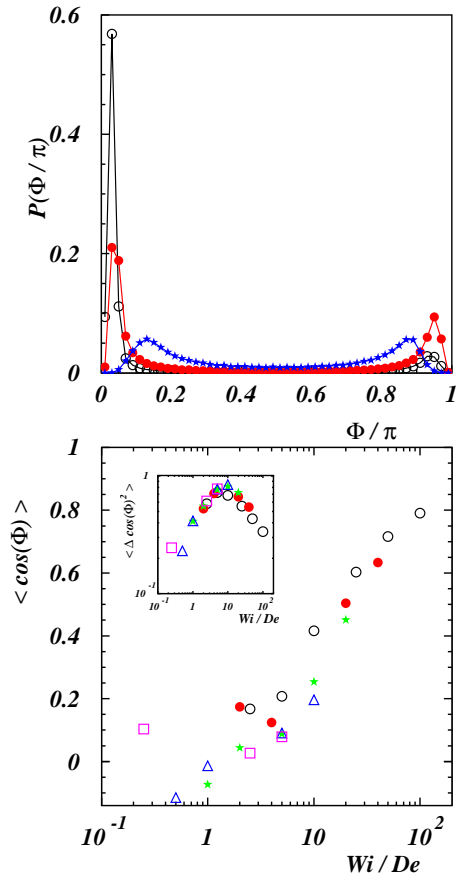}
\caption{(Top) Probability distribution function
of the polymer inclination angle $\Phi$ based on data from positive half-cycle 
for $L_p/L=0.5$, $De=10$ ($\circ$, black), $50$ ($\bullet$, red), 
$200$ ($\star$, blue), and $Wi=500$. The distribution functions for 
negative half-cycle are mirror symmetric with respect to $\Phi/\pi =1/2$.
(Bottom) Average of $\cos \Phi$ 
as a function of the strain $Wi/De$
for $L_p/L=0.5$
and $De=10$ ($\circ$, black), $25$ ($\bullet$, red), $50$ ($\star$, green), 
$100$ ($\triangle$, blue), and $200$ ($\square$, magenta). 
The inset shows the variance 
$\langle(\cos \Phi - \langle \cos \Phi\rangle)^2 \rangle$ 
as a function of $Wi/De$ for
$L_p/L=0.5$ and $De=10$ ($\circ$, black), $25$ ($\bullet$, red), 
$50$ ($\star$, green), 
$100$ ($\triangle$, blue), and $200$ ($\square$, magenta).
}
\label{fig:pdfangle_incl}
\end{center}
\end{figure}

The described mechanism of chain flipping is supported by the probability distribution functions $P(\Phi)$ of the inclination angle $\Phi$, 
which is
defined as the angle between the polymer center-of-mass position vector and the flow direction (Fig.~\ref{fig:pdfangle_incl}). 
\begin{figure*}[ht] 
\begin{center}
\includegraphics*[width=0.8\textwidth,angle=0]{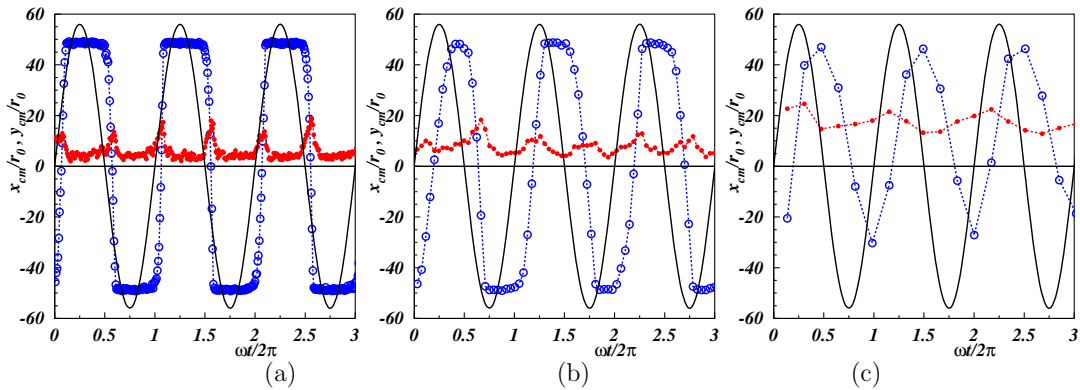}
\caption{Time dependence of the  polymer center-of-mass position along the flow ($\circ$, blue), $x_{cm}$,
and the gradient ($\bullet$, red), $y_{cm}$, directions for $L_p/L=0.5$, 
$Wi=500$, and $De=10$(a), $50$(b), and $200$(c).
The black lines indicate the externally applied shear flow with arbitrary amplitude, and the dashed lines are guides for the eye.
\label{fig:cmvst}
}
\end{center}
\end{figure*}
For large strain, the polymers spend most of their time aligned with the flow direction (the main peak is located
at $\Phi < \pi/2$) during the positive half-cycle of the shear oscillation
(Fig.~\ref{fig:pdfangle_incl} (left)), and opposite to the flow
direction (the main peak is at $\Phi > \pi/2$)
during the negative half-cycle of the shear oscillation
(the distribution functions for the negative half-cycle are mirror symmetric with respect to $\Phi=\pi/2$). 
This suggests that for high values of the
strain the inclination angle and the shear flow are of the same sign before
the shear flow changes sign. 
Interestingly, for the lowest considered value $Wi/De=2.5$, the distribution function $P(\Phi)$ for the positive and negative half-cycles are indistinguishable for the considered  persistence lengths---$P(\Phi)$
shows two broad peaks of comparable height symmetric with respect to the
shear direction ($\Phi = \pi/2$). This indicates that polymers
can either flip or keep their orientation with equal probability, when the shear rate changes sign.

The average, $\langle \cos \Phi \rangle$ (Fig.~\ref{fig:pdfangle_incl} (right)), 
of the inclination angle depends weakly on shear and increases logarithmically,  $\langle \cos \Phi \rangle \sim \log (Wi/De)$, on strain,
while its variance has a maximum for intermediate values of strain.

\subsubsection{Center-of-mass motion}

\begin{figure}[ht]
\begin{center}
\includegraphics*[width=0.7\columnwidth,angle=0]{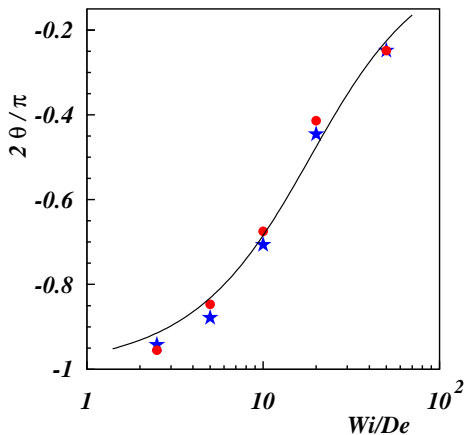}
\caption{Dependence of the phase shift $\theta$ on the strain $Wi/De$
for $L_p/L=0.5$ ($\star$, blue) and $2$ ($\bullet$, red), with
$Wi=500$. The full line represents the fitted functional dependence 
$2 \arccot\left (-0.054 Wi/De \right)/\pi$.
\label{fig:phaseshift}
}
\end{center}
\end{figure}

Figure~\ref{fig:cmvst} illustrates the time dependence of the center-of-mass, $\bm r_{cm} = (x_{cm}, y_{cm})^T$,  during three 
periods for $L_p/L=0.5$. 
For the
highest considered strain $Wi/De=50$ (Fig.~\ref{fig:cmvst}(a)),
$x_{cm}$ exhibits a nearly in-phase periodic motion with a small
phase shift $\theta$, and the dynamics is apparently no longer harmonic.
The component $y_{cm}$ shows very narrow peaks whenever $x_{cm}$ changes sign and becomes very small as soon as the polymer is stretched. Remarkably,  $y_{cm}$ exhibits frequency doubling. 
We like to stress that the frequency doubling is independent of the considered quantity used to characterize the center-of-mass motion. Specifically, we considered the time dependence of the inclination angle $\Phi$ and
that of the distance $R_{cm}=\sqrt{x_{cm}^2+y_{cm}^2}$. While $\Phi$ exhibits the same
frequency as the applied flow, $R_{cm}$ shows frequency doubling.
With decreasing $Wi/De$, the motion is still periodic and
becomes more harmonic with a larger phase shift (Fig.~\ref{fig:cmvst}(b)).
The $y_{cm}$ peaks are now broader and their amplitude decreases.
Finally, for the smallest value $Wi/De=2.5$ (Fig.~\ref{fig:cmvst}(c)),
$x_{cm}$ partially follows the external flow with a reduced amplitude
which is no longer symmetric with respect to  $x=0$ due to entropic effects, and  the magnitude of the angle $\theta$ further
increases. The time sequence of $y_{cm}$ indicates that the polymer is no longer fully elongated
along the wall when the flow reaches its extreme values. The peaks are very
broad and no clear indication of a frequency doubling can be
observed, since the amplitude of $y_{cm}$ is significantly reduced.
The comparable data for $L_p/L=2$ show no appreciable differences.
In order to evaluate the phase shift $\theta$, the values of
$x_{cm}$ are fitted to the function
$A \sin(\omega t + \theta)$, where $A$ and $\theta$ are fitting parameters.
The values of $\theta$ are in the range 
$(-\pi/2, 0)$  
and appear to be independent
on the persistence length (Fig.~\ref{fig:phaseshift}). The extracted 
dependence on $Wi/De$ can well be described by  the function
$\theta = \arccot\left(-a Wi/De \right)$, with $a=0.054$.

\begin{figure}[ht]
\begin{center}
\includegraphics*[width=\columnwidth,angle=0]{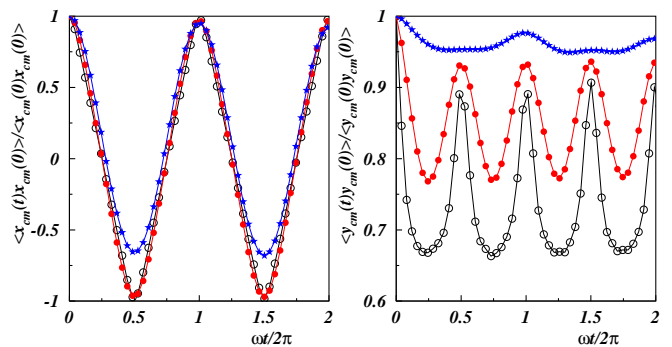}
\caption{Time auto-correlations of the center-of-mass $x_{cm}$ (left) and
$y_{cm}$ (right) Cartesian coordinates
for $L_p/L=2$, $Wi=500$, and $De=10$ ($\circ$, black), $50$ ($\bullet$, red),
$200$ ($\star$, blue).
\label{fig:autocorrcm}
}
\end{center}
\end{figure}

Further evidence of the discussed frequency doubling is obtained by the autocorrelation functions of the  Cartesian center-of-mass coordinates (Fig.~\ref{fig:autocorrcm}). Consistent with Fig.~\ref{fig:cmvst}, the correlation functions display a periodic motion.  The $x_{cm}$-component reveals the same frequency as the external flow.
More importantly, the $y_{cm}$-component clearly shows
frequency doubling for $Wi/De \gtrsim 10$.
The wall at $y=0$ reflects the polymer  every time the flow moves the polymer
from one side to the other. This process occurs twice in every flow cycle
causing the observed frequency doubling in the dynamics along the $y$ axis.

\subsection{Internal polymer dynamics} \label{subsec:modes}

\begin{figure}[ht]
\begin{center}
\includegraphics*[width=0.7\columnwidth,angle=0]{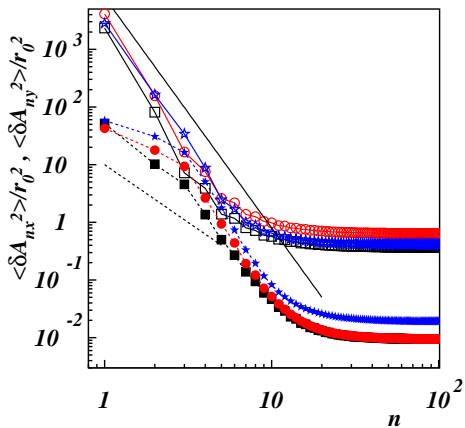}
\caption{Variance of the mode amplitudes $A_{x,n}$ (empty symbols connected by full lines)
and $A_{y,n}$ (filled symbols connected by dashed lines)
as functions of mode number $n$ for $L_p/L=2$ with
$Wi=500$ and $De=10$ ($\square$, black), $50$ ($\circ$, red), $200$ ($\star$, blue).
The slopes of full and dashed lines are $-4$ and $-2$, respectively.
\label{fig:undul}
}
\end{center}
\end{figure}

The snapshots of Fig.~\ref{fig:conf05} reveal strong polymer conformational changes during oscillation cycles. We characterize the appearing internal polymer dynamics by the normal mode expansion 
\begin{equation} \label{modes}
{\bm r}_i = \sum_{n=1}^{N-1}{\bm A}_n(t) \sin[q_n(i-1)] \; , \;\;\; i=1,\ldots,N ,
\end{equation}
with the wave vectors  $q_n=(n-1/2) \pi/(N-1)$ ($n=1,\ldots,N-1$).  The normal mode amplitudes $\bm A_n(t) = (A_{x,n}(t), A_{y,n}(t))^T$ are given by
\begin{eqnarray}\label{comp}
A_{x,n}&=&\frac{2}{N-1}\sum_{i=1}^{N} r_{x,i} \sin[q_n(i-1)] ,\\
A_{y,n}&=&\frac{2}{N-1}\sum_{i=1}^{N} r_{y,i} \sin[q_n(i-1)] .
\end{eqnarray}
The dependence of the variance of the mode amplitudes
$\langle \delta \bm A_n^2 \rangle =
\langle \left( \bm A_n- \langle \bm A_n \rangle \right)^2 \rangle$
on the mode number is illustrated in Fig.~\ref{fig:undul} for the stiffer polymer. 
Evidently, the variances of $A_{x,n}$ and $A_{y,n}$ are
distinctly different for the considered strains. In the absence of shear, both components of the mode amplitudes exhibit the mode number dependence $n^{-4}$, in accordance with the semiflexible character of the considered polymers \cite{harn:95,arag:85}.  Note that the plateau for large mode numbers is a consequence of the discreteness of the polymer with a finite number of modes. 
Along the flow direction, the  $n^{-4}$ dependence persists even under shear due to the strong polymer stretching. However, we observe a pronounced strain effects on $A_{y,n}$ for small mode numbers $3 \leq n \leq 5$. The indicated dependence $n^{-2}$, typical for flexible polymers \cite{harn:95,rous:53,doi:86}, suggests that the polymer acquires a flexible polymer-like behavior on larger length scales, with a crossover to semiflexibility,  $n^{-4}$ dependence, on smaller scale,  along the gradient direction. A similar behavior is found for the more flexible chain.

\begin{figure}[ht]
\begin{center}
\includegraphics*[width=\columnwidth,angle=0]{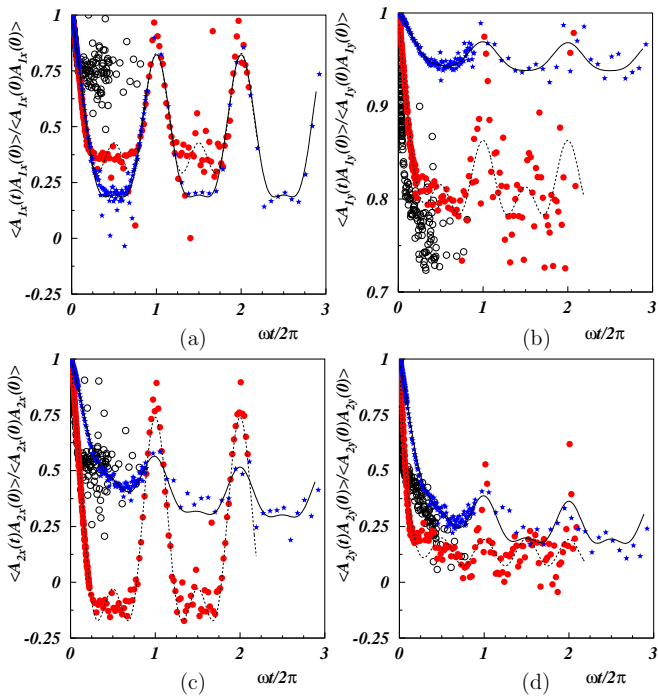}
\caption{Autocorrelation function of the mode amplitudes
for the modes $n=1$ along  the (a) $x$- and (b) $y$-direction as well as for $n=2$ along the (c) $x$- and (d) $y$-direction as a function 
of  time with $L_p/L=2$, $Wi=500$, and
$De=10$ ($\circ$, black), $50$ ($\bullet$, red), $200$ ($\star$, blue). 
The black dotted and solid lines are fits of Eq.~(\eqref{eq:fit}) to the data for $De=50$ and  $De=200$, respectively.
\label{fig:autocorr}
}
\end{center}
\end{figure}

The  dynamics  of  the  mode  amplitudes  with $n=1, 2$ is analyzed in terms of the mode-autocorrelation functions
$\lla A_{x,n}(t)A_{x,n}(0) \rra$ and $\lla A_{y,n}(t)A_{y,n}(0) \rra$.
Results for various strains and
the persistence length $L_p/L=2$ are presented in Fig.~\ref{fig:autocorr}.
The simulation data are fitted by the function 
\begin{align} \label{eq:fit} \nonumber 
f(t)= & \ A \left[ \exp(-\gamma t/T) -1 \right]
+ B\left[ \cos(\Omega \omega t) -1 \right] \\ &
 + C \left[ \cos(2 \Omega \omega t) -1 \right] +1,
\end{align}
where the dimensionless factors $\gamma$ and $\Omega$ characterizes the relaxation time 
and  possible variations of the frequency
with respect to the imposed  $\omega$, respectively.
The correlation functions for $De=50$ and $De=200$ are well fitted by 
Eq.~(\ref{eq:fit}), while  data for
$De=10$ do not allow for a suitable fitting. 
The factor $\gamma$ of the $x$ component decreases with increasing mode number, $n$,
revealing a faster  relaxation on larger length-scales along the flow direction.  Thereby,  $\gamma$ depends on $De$ and is  larger for $De=200$. It  increases with increasing  ratio $Wi/De$, reflecting the strong dependence of the relaxation process on the external field. 
 
At equilibrium, the variance $\langle \delta \bm A_n^2 \rangle $ 
of the amplitudes  is determined by the polymer relaxation times \cite{doi:86}. 
The mode-number dependence of $\gamma$  is inconsistent with that of  $\langle \delta \bm A_n^2 \rangle $  (Fig.~\ref{fig:undul}),  which indicates a strong influence of the external field on the time dependence of the internal relaxation process.  Along  the $y$ direction, $\gamma$  is approximately constant at fixed strain, indicating that the chain relaxation time does not vary significantly on large scales along the shear direction.

The  factor  $\Omega$ is close to unity and oscillations occur with  the external frequency. For both, the $x$ and $y$ components, 
the  term with $2\omega$ contributes to the oscillations of the correlation function, i.e., the observed  frequency doubling  found for the center-of-mass motion is also reflected in the internal dynamics.

\section{Conclusions} \label{sec:conclusions}

We have analyzed the conformational and dynamical properties of semiflexible polymers tethered at an impenetrable wall under oscillatory shear flow.  We identify three  different regimes in terms of polymer deformation as a function of strain. At small strain, $Wi/De \lesssim 1$, the polymer structures are close to the equilibrium conformations and they are hardly affected by flow. For intermediate strain, $1 < Wi/De < 10$, 
the polymer is gradually stretched and its end-to-end distance increases logarithmically. Above $Wi/De \gtrsim 10$, the deficit length exhibits a power-law decrease with increasing strain, in agreement with experimental results under steady shear flow \cite{doyl:00,lado:00}. 

The polymer conformations are tightly coupled with their dynamics. At high strain, the polymers follow the external oscillations of the flow with a small phase shift, and  are strongly stretched and aligned with the wall. The polymer dynamics is far less enslaved by the flow field for intermediate strains, which is reflected by a larger phase difference between the flow-induced polymer structures and the external flow. Most importantly, this allows for more swollen polymer-like conformations and a higher polymer density above the wall. 

As might be expected for the considered geometry, conformational properties normal to the wall exhibit frequency doubling compared to the external frequency $\omega$, for example, the autocorrelation function of the $y$-center-of-mass coordinate. While the properties along the  $x$ direction are typically symmetric in terms of positive and negative values, those along $y$ are not and only positive values are possible. Hence, positive $y$ values correspond to positive and negative $x$ values, which determines the frequency doubling.    

In absence of chirality and torsion, we expect semiflexible polymers in three dimensions to exhibit similar features 
in terms of conformational changes. 
This is reflected in the similarity of our results 
with experiments of DNA molecules tethered to a wall under shear flow \cite{lado:00}.  
Hydrodynamic interactions influence the polymer dynamics, however, 
the extent depends on their stiffness.  
As far as semiflexible polymers
are concerned, hydrodynamic interactions are of minor importance, so that  
the observed conformational
and dynamical features will also be present for polymers
embedded in a fluid even in three dimensions.
Hydrodynamic interactions are expected to more severely
affect the dynamics of flexible polymers. They  are again weak for rather stretched conformations \cite{lado:00}, 
but can dominate the dynamics in the coiled state and  accelerate the polymer motion.
In particular,  this affects the internal dynamics along the $y$ direction, with a rather Zimm- than Rouse-like  
mode dependence of the relaxation times on larger length scales \cite{doi:86}. 

Our simulations provide valuable insight into the properties of polymers exposed to oscillatory flows. Specifically, we demonstrate that such simulations are feasible for experimentally relevant parameters. Our studies are a first step only toward systematic investigations of polymer properties under oscillatory shear and we hope that they will prompt further simulations along that line. 

\section*{Supplementary material}

The supplementary movies show the animations of the polymer dynamics 
over two shear periods for the cases reported in Fig.~\ref{fig:conf05}.
Movie 1 is relative to the case with $L_p/L=0.5$, $Wi=500$, $De=10$,
movie 2 is relative to the case with $L_p/L=2$, $Wi=500$, $De=10$,
movie 3 is relative to the case with $L_p/L=0.5$, $Wi=500$, $De=200$, and
movie 4 is relative to the case with $L_p/L=2$, $Wi=500$, $De=200$.

\acknowledgments
A. L. thanks G. G., R. G. W.,  and co-workers for their hospitality during the stay at
the Forschungszentrum J\"{u}lich and acknowledges support from the DAAD through the 
``Research Stays for University Academics and Scientists'' program.

\section*{Data Availability Statement}
The data that support the findings of this study are available 
from the corresponding author upon reasonable request.


\begin{thebibliography}{70}%
\makeatletter
\providecommand \@ifxundefined [1]{%
 \@ifx{#1\undefined}
}%
\providecommand \@ifnum [1]{%
 \ifnum #1\expandafter \@firstoftwo
 \else \expandafter \@secondoftwo
 \fi
}%
\providecommand \@ifx [1]{%
 \ifx #1\expandafter \@firstoftwo
 \else \expandafter \@secondoftwo
 \fi
}%
\providecommand \natexlab [1]{#1}%
\providecommand \enquote  [1]{``#1''}%
\providecommand \bibnamefont  [1]{#1}%
\providecommand \bibfnamefont [1]{#1}%
\providecommand \citenamefont [1]{#1}%
\providecommand \href@noop [0]{\@secondoftwo}%
\providecommand \href [0]{\begingroup \@sanitize@url \@href}%
\providecommand \@href[1]{\@@startlink{#1}\@@href}%
\providecommand \@@href[1]{\endgroup#1\@@endlink}%
\providecommand \@sanitize@url [0]{\catcode `\\12\catcode `\$12\catcode
  `\&12\catcode `\#12\catcode `\^12\catcode `\_12\catcode `\%12\relax}%
\providecommand \@@startlink[1]{}%
\providecommand \@@endlink[0]{}%
\providecommand \url  [0]{\begingroup\@sanitize@url \@url }%
\providecommand \@url [1]{\endgroup\@href {#1}{\urlprefix }}%
\providecommand \urlprefix  [0]{URL }%
\providecommand \Eprint [0]{\href }%
\providecommand \doibase [0]{http://dx.doi.org/}%
\providecommand \selectlanguage [0]{\@gobble}%
\providecommand \bibinfo  [0]{\@secondoftwo}%
\providecommand \bibfield  [0]{\@secondoftwo}%
\providecommand \translation [1]{[#1]}%
\providecommand \BibitemOpen [0]{}%
\providecommand \bibitemStop [0]{}%
\providecommand \bibitemNoStop [0]{.\EOS\space}%
\providecommand \EOS [0]{\spacefactor3000\relax}%
\providecommand \BibitemShut  [1]{\csname bibitem#1\endcsname}%
\let\auto@bib@innerbib\@empty
\bibitem [{\citenamefont {Bird}, \citenamefont {Armstrong},\ and\ \citenamefont
  {Hassager}(1987)}]{bird:87.1}%
  \BibitemOpen
  \bibfield  {author} {\bibinfo {author} {\bibfnamefont {R.~B.}\ \bibnamefont
  {Bird}}, \bibinfo {author} {\bibfnamefont {R.~C.}\ \bibnamefont {Armstrong}},
  \ and\ \bibinfo {author} {\bibfnamefont {O.}~\bibnamefont {Hassager}},\
  }\href@noop {} {\emph {\bibinfo {title} {Dynamics of Polymer Liquids}}},\
  Vol.~\bibinfo {volume} {1}\ (\bibinfo  {publisher} {John Wiley \& Sons},\
  \bibinfo {address} {New York},\ \bibinfo {year} {1987})\BibitemShut {NoStop}%
\bibitem [{\citenamefont {Larson}(1999)}]{lars:99}%
  \BibitemOpen
  \bibfield  {author} {\bibinfo {author} {\bibfnamefont {R.~G.}\ \bibnamefont
  {Larson}},\ }\href@noop {} {\emph {\bibinfo {title} {The Structure and
  Rheology of Complex Fluids}}}\ (\bibinfo  {publisher} {Oxford University
  Press},\ \bibinfo {address} {New York},\ \bibinfo {year} {1999})\BibitemShut
  {NoStop}%
\bibitem [{\citenamefont {Rubinstein}\ and\ \citenamefont
  {Colby}(2003)}]{rubi:03}%
  \BibitemOpen
  \bibfield  {author} {\bibinfo {author} {\bibfnamefont {M.}~\bibnamefont
  {Rubinstein}}\ and\ \bibinfo {author} {\bibfnamefont {R.~C.}\ \bibnamefont
  {Colby}},\ }\href@noop {} {\emph {\bibinfo {title} {Polymer Physics}}}\
  (\bibinfo  {publisher} {Oxford University Press, Oxford},\ \bibinfo {year}
  {2003})\BibitemShut {NoStop}%
\bibitem [{\citenamefont {Larson}\ and\ \citenamefont {Desai}(2015)}]{lars:15}%
  \BibitemOpen
  \bibfield  {author} {\bibinfo {author} {\bibfnamefont {R.~G.}\ \bibnamefont
  {Larson}}\ and\ \bibinfo {author} {\bibfnamefont {P.~S.}\ \bibnamefont
  {Desai}},\ }\bibfield  {title} {\enquote {\bibinfo {title} {Modeling the
  rheology of polymer melts and solutions},}\ }\href@noop {} {\bibfield
  {journal} {\bibinfo  {journal} {Ann. Rev. Fluid Mech.}\ }\textbf {\bibinfo
  {volume} {47}},\ \bibinfo {pages} {47} (\bibinfo {year} {2015})}\BibitemShut
  {NoStop}%
\bibitem [{\citenamefont {Shaw}\ and\ \citenamefont
  {MacKnight}(2018)}]{shaw:18}%
  \BibitemOpen
  \bibfield  {author} {\bibinfo {author} {\bibfnamefont {M.~T.}\ \bibnamefont
  {Shaw}}\ and\ \bibinfo {author} {\bibfnamefont {W.~J.}\ \bibnamefont
  {MacKnight}},\ }\href@noop {} {\emph {\bibinfo {title} {Introduction to
  polymer viscoelasticity}}}\ (\bibinfo  {publisher} {John Wiley \& Sons},\
  \bibinfo {year} {2018})\BibitemShut {NoStop}%
\bibitem [{\citenamefont {Schroeder}(2018)}]{schr:18}%
  \BibitemOpen
  \bibfield  {author} {\bibinfo {author} {\bibfnamefont {C.~M.}\ \bibnamefont
  {Schroeder}},\ }\href@noop {} {\bibfield  {journal} {\bibinfo  {journal} {J.
  Rheol.}\ }\textbf {\bibinfo {volume} {62}},\ \bibinfo {pages} {371} (\bibinfo
  {year} {2018})}\BibitemShut {NoStop}%
\bibitem [{\citenamefont {Perkins}\ \emph {et~al.}(1995)\citenamefont
  {Perkins}, \citenamefont {Smith}, \citenamefont {Larson},\ and\ \citenamefont
  {Chu}}]{perk:95}%
  \BibitemOpen
  \bibfield  {author} {\bibinfo {author} {\bibfnamefont {T.~T.}\ \bibnamefont
  {Perkins}}, \bibinfo {author} {\bibfnamefont {D.~E.}\ \bibnamefont {Smith}},
  \bibinfo {author} {\bibfnamefont {R.~G.}\ \bibnamefont {Larson}}, \ and\
  \bibinfo {author} {\bibfnamefont {S.}~\bibnamefont {Chu}},\ }\href@noop {}
  {\bibfield  {journal} {\bibinfo  {journal} {Science}\ }\textbf {\bibinfo
  {volume} {268}},\ \bibinfo {pages} {83} (\bibinfo {year} {1995})}\BibitemShut
  {NoStop}%
\bibitem [{\citenamefont {Winkler}(2006{\natexlab{a}})}]{wink:06.1}%
  \BibitemOpen
  \bibfield  {author} {\bibinfo {author} {\bibfnamefont {R.~G.}\ \bibnamefont
  {Winkler}},\ }\bibfield  {title} {\enquote {\bibinfo {title} {Semiflexible
  polymers in shear flow},}\ }\href {\doibase 10.1103/PhysRevLett.97.128301}
  {\bibfield  {journal} {\bibinfo  {journal} {Phys. Rev. Lett.}\ }\textbf
  {\bibinfo {volume} {97}},\ \bibinfo {pages} {128301} (\bibinfo {year}
  {2006}{\natexlab{a}})}\BibitemShut {NoStop}%
\bibitem [{\citenamefont {Winkler}(2010)}]{wink:10}%
  \BibitemOpen
  \bibfield  {author} {\bibinfo {author} {\bibfnamefont {R.~G.}\ \bibnamefont
  {Winkler}},\ }\bibfield  {title} {\enquote {\bibinfo {title} {Conformational
  and rheological properties of semiflexible polymers in shear flow},}\
  }\href@noop {} {\bibfield  {journal} {\bibinfo  {journal} {J. Chem. Phys.}\
  }\textbf {\bibinfo {volume} {133}},\ \bibinfo {pages} {164905} (\bibinfo
  {year} {2010})}\BibitemShut {NoStop}%
\bibitem [{\citenamefont {Wilhelm}\ and\ \citenamefont {Frey}(1996)}]{wilh:96}%
  \BibitemOpen
  \bibfield  {author} {\bibinfo {author} {\bibfnamefont {J.}~\bibnamefont
  {Wilhelm}}\ and\ \bibinfo {author} {\bibfnamefont {E.}~\bibnamefont {Frey}},\
  }\bibfield  {title} {\enquote {\bibinfo {title} {Radial distribution function
  of semiflexible polymers},}\ }\href@noop {} {\bibfield  {journal} {\bibinfo
  {journal} {Phys. Rev. Lett.}\ }\textbf {\bibinfo {volume} {77}},\ \bibinfo
  {pages} {2581} (\bibinfo {year} {1996})}\BibitemShut {NoStop}%
\bibitem [{\citenamefont {G{\"o}tter}\ \emph {et~al.}(1996)\citenamefont
  {G{\"o}tter}, \citenamefont {Kroy}, \citenamefont {Frey}, \citenamefont
  {B{\"a}rmann},\ and\ \citenamefont {Sackmann}}]{goet:96}%
  \BibitemOpen
  \bibfield  {author} {\bibinfo {author} {\bibfnamefont {R.}~\bibnamefont
  {G{\"o}tter}}, \bibinfo {author} {\bibfnamefont {K.}~\bibnamefont {Kroy}},
  \bibinfo {author} {\bibfnamefont {E.}~\bibnamefont {Frey}}, \bibinfo {author}
  {\bibfnamefont {M.}~\bibnamefont {B{\"a}rmann}}, \ and\ \bibinfo {author}
  {\bibfnamefont {E.}~\bibnamefont {Sackmann}},\ }\href@noop {} {\bibfield
  {journal} {\bibinfo  {journal} {Macromolecules}\ }\textbf {\bibinfo {volume}
  {29}},\ \bibinfo {pages} {30} (\bibinfo {year} {1996})}\BibitemShut {NoStop}%
\bibitem [{\citenamefont {Harnau}, \citenamefont {Winkler},\ and\ \citenamefont
  {Reineker}(1996)}]{harn:96}%
  \BibitemOpen
  \bibfield  {author} {\bibinfo {author} {\bibfnamefont {L.}~\bibnamefont
  {Harnau}}, \bibinfo {author} {\bibfnamefont {R.~G.}\ \bibnamefont {Winkler}},
  \ and\ \bibinfo {author} {\bibfnamefont {P.}~\bibnamefont {Reineker}},\
  }\bibfield  {title} {\enquote {\bibinfo {title} {Dynamic structure factor of
  semiflexible macromolecules in dilute solution},}\ }\href@noop {} {\bibfield
  {journal} {\bibinfo  {journal} {J. Chem. Phys.}\ }\textbf {\bibinfo {volume}
  {104}},\ \bibinfo {pages} {6355} (\bibinfo {year} {1996})}\BibitemShut
  {NoStop}%
\bibitem [{\citenamefont {Everaers}\ \emph {et~al.}(1999)\citenamefont
  {Everaers}, \citenamefont {J\"ulicher}, \citenamefont {Ajdari},\ and\
  \citenamefont {Maggs}}]{ever:99}%
  \BibitemOpen
  \bibfield  {author} {\bibinfo {author} {\bibfnamefont {R.}~\bibnamefont
  {Everaers}}, \bibinfo {author} {\bibfnamefont {F.}~\bibnamefont
  {J\"ulicher}}, \bibinfo {author} {\bibfnamefont {A.}~\bibnamefont {Ajdari}},
  \ and\ \bibinfo {author} {\bibfnamefont {A.~C.}\ \bibnamefont {Maggs}},\
  }\bibfield  {title} {\enquote {\bibinfo {title} {Dynamic fluctuations of
  semiflexible filaments},}\ }\href@noop {} {\bibfield  {journal} {\bibinfo
  {journal} {Phys. Rev. Lett.}\ }\textbf {\bibinfo {volume} {82}},\ \bibinfo
  {pages} {3717} (\bibinfo {year} {1999})}\BibitemShut {NoStop}%
\bibitem [{\citenamefont {Samuel}\ and\ \citenamefont {Sinha}(2002)}]{samu:02}%
  \BibitemOpen
  \bibfield  {author} {\bibinfo {author} {\bibfnamefont {J.}~\bibnamefont
  {Samuel}}\ and\ \bibinfo {author} {\bibfnamefont {S.}~\bibnamefont {Sinha}},\
  }\bibfield  {title} {\enquote {\bibinfo {title} {Elasticity of semiflexible
  polymers},}\ }\href@noop {} {\bibfield  {journal} {\bibinfo  {journal} {Phys.
  Rev. E}\ }\textbf {\bibinfo {volume} {66}},\ \bibinfo {pages} {050801}
  (\bibinfo {year} {2002})}\BibitemShut {NoStop}%
\bibitem [{\citenamefont {Le~Goff}\ \emph {et~al.}(2002)\citenamefont
  {Le~Goff}, \citenamefont {Hallatschek}, \citenamefont {Frey},\ and\
  \citenamefont {Amblard}}]{lego:02}%
  \BibitemOpen
  \bibfield  {author} {\bibinfo {author} {\bibfnamefont {L.}~\bibnamefont
  {Le~Goff}}, \bibinfo {author} {\bibfnamefont {O.}~\bibnamefont
  {Hallatschek}}, \bibinfo {author} {\bibfnamefont {E.}~\bibnamefont {Frey}}, \
  and\ \bibinfo {author} {\bibfnamefont {F.}~\bibnamefont {Amblard}},\
  }\bibfield  {title} {\enquote {\bibinfo {title} {Tracer studies on f-actin
  fluctuations},}\ }\href@noop {} {\bibfield  {journal} {\bibinfo  {journal}
  {Phys. Rev. Lett.}\ }\textbf {\bibinfo {volume} {89}},\ \bibinfo {pages}
  {258101} (\bibinfo {year} {2002})}\BibitemShut {NoStop}%
\bibitem [{\citenamefont {Winkler}(2003)}]{wink:03}%
  \BibitemOpen
  \bibfield  {author} {\bibinfo {author} {\bibfnamefont {R.~G.}\ \bibnamefont
  {Winkler}},\ }\bibfield  {title} {\enquote {\bibinfo {title} {Deformation of
  semiflexible chains},}\ }\href@noop {} {\bibfield  {journal} {\bibinfo
  {journal} {J. Chem. Phys.}\ }\textbf {\bibinfo {volume} {118}},\ \bibinfo
  {pages} {2919} (\bibinfo {year} {2003})}\BibitemShut {NoStop}%
\bibitem [{\citenamefont {Petrov}\ \emph {et~al.}(2006)\citenamefont {Petrov},
  \citenamefont {Ohrt}, \citenamefont {Winkler},\ and\ \citenamefont
  {Schwille}}]{petr:06}%
  \BibitemOpen
  \bibfield  {author} {\bibinfo {author} {\bibfnamefont {E.~P.}\ \bibnamefont
  {Petrov}}, \bibinfo {author} {\bibfnamefont {T.}~\bibnamefont {Ohrt}},
  \bibinfo {author} {\bibfnamefont {R.~G.}\ \bibnamefont {Winkler}}, \ and\
  \bibinfo {author} {\bibfnamefont {P.}~\bibnamefont {Schwille}},\ }\bibfield
  {title} {\enquote {\bibinfo {title} {Diffusion and segmental dynamics of
  double-stranded {DNA}},}\ }\href@noop {} {\bibfield  {journal} {\bibinfo
  {journal} {Phys. Rev. Lett.}\ }\textbf {\bibinfo {volume} {97}},\ \bibinfo
  {pages} {258101} (\bibinfo {year} {2006})}\BibitemShut {NoStop}%
\bibitem [{\citenamefont {Bird}\ \emph {et~al.}(1987)\citenamefont {Bird},
  \citenamefont {Curtiss}, \citenamefont {Armstrong},\ and\ \citenamefont
  {Hassager}}]{bird:87}%
  \BibitemOpen
  \bibfield  {author} {\bibinfo {author} {\bibfnamefont {R.~B.}\ \bibnamefont
  {Bird}}, \bibinfo {author} {\bibfnamefont {C.~F.}\ \bibnamefont {Curtiss}},
  \bibinfo {author} {\bibfnamefont {R.~C.}\ \bibnamefont {Armstrong}}, \ and\
  \bibinfo {author} {\bibfnamefont {O.}~\bibnamefont {Hassager}},\ }\href@noop
  {} {\emph {\bibinfo {title} {Dynamics of Polymer Liquids}}},\ Vol.~\bibinfo
  {volume} {2}\ (\bibinfo  {publisher} {John Wiley \& Sons},\ \bibinfo
  {address} {New York},\ \bibinfo {year} {1987})\BibitemShut {NoStop}%
\bibitem [{\citenamefont {{\"O}ttinger}(1996)}]{oett:96}%
  \BibitemOpen
  \bibfield  {author} {\bibinfo {author} {\bibfnamefont {H.~C.}\ \bibnamefont
  {{\"O}ttinger}},\ }\href@noop {} {\emph {\bibinfo {title} {Stochastic
  Processes in Polymeric Fluids}}}\ (\bibinfo  {publisher} {Springer},\
  \bibinfo {address} {Berlin},\ \bibinfo {year} {1996})\BibitemShut {NoStop}%
\bibitem [{\citenamefont {Winkler}(2006{\natexlab{b}})}]{wink:06_1}%
  \BibitemOpen
  \bibfield  {author} {\bibinfo {author} {\bibfnamefont {R.~G.}\ \bibnamefont
  {Winkler}},\ }\bibfield  {title} {\enquote {\bibinfo {title} {Semiflexible
  polymers in shear flow},}\ }\href@noop {} {\bibfield  {journal} {\bibinfo
  {journal} {Phys. Rev. Lett.}\ }\textbf {\bibinfo {volume} {97}},\ \bibinfo
  {pages} {128301} (\bibinfo {year} {2006}{\natexlab{b}})}\BibitemShut
  {NoStop}%
\bibitem [{\citenamefont {Munk}\ \emph {et~al.}(2006)\citenamefont {Munk},
  \citenamefont {Hallatschek}, \citenamefont {Wiggins},\ and\ \citenamefont
  {Frey}}]{munk:06}%
  \BibitemOpen
  \bibfield  {author} {\bibinfo {author} {\bibfnamefont {T.}~\bibnamefont
  {Munk}}, \bibinfo {author} {\bibfnamefont {O.}~\bibnamefont {Hallatschek}},
  \bibinfo {author} {\bibfnamefont {C.~H.}\ \bibnamefont {Wiggins}}, \ and\
  \bibinfo {author} {\bibfnamefont {E.}~\bibnamefont {Frey}},\ }\bibfield
  {title} {\enquote {\bibinfo {title} {Dynamics of semiflexible polymers in a
  flow field},}\ }\href@noop {} {\bibfield  {journal} {\bibinfo  {journal}
  {Phys. Rev. E}\ }\textbf {\bibinfo {volume} {74}},\ \bibinfo {pages} {041911}
  (\bibinfo {year} {2006})}\BibitemShut {NoStop}%
\bibitem [{\citenamefont {Hur}\ and\ \citenamefont {Shaqfeh}(2000)}]{hur:00}%
  \BibitemOpen
  \bibfield  {author} {\bibinfo {author} {\bibfnamefont {J.~S.}\ \bibnamefont
  {Hur}}\ and\ \bibinfo {author} {\bibfnamefont {E.~S.~G.}\ \bibnamefont
  {Shaqfeh}},\ }\bibfield  {title} {\enquote {\bibinfo {title} {Brownian
  dynamics simulations of single {DNA} molecules in shear flow},}\ }\href@noop
  {} {\bibfield  {journal} {\bibinfo  {journal} {J. Rheol.}\ }\textbf {\bibinfo
  {volume} {44}},\ \bibinfo {pages} {713} (\bibinfo {year} {2000})}\BibitemShut
  {NoStop}%
\bibitem [{\citenamefont {Jendrejack}, \citenamefont {de~Pablo},\ and\
  \citenamefont {Graham}(2002)}]{jend:02}%
  \BibitemOpen
  \bibfield  {author} {\bibinfo {author} {\bibfnamefont {R.~M.}\ \bibnamefont
  {Jendrejack}}, \bibinfo {author} {\bibfnamefont {J.~J.}\ \bibnamefont
  {de~Pablo}}, \ and\ \bibinfo {author} {\bibfnamefont {M.~D.}\ \bibnamefont
  {Graham}},\ }\bibfield  {title} {\enquote {\bibinfo {title} {Stochastic
  simulations of dna in flow: Dynamics and the effects of hydrodynamic
  interactions},}\ }\href@noop {} {\bibfield  {journal} {\bibinfo  {journal}
  {J. Chem. Phys.}\ }\textbf {\bibinfo {volume} {116}},\ \bibinfo {pages}
  {7752} (\bibinfo {year} {2002})}\BibitemShut {NoStop}%
\bibitem [{\citenamefont {Hsieh}\ and\ \citenamefont {Larson}(2004)}]{hsie:04}%
  \BibitemOpen
  \bibfield  {author} {\bibinfo {author} {\bibfnamefont {C.-C.}\ \bibnamefont
  {Hsieh}}\ and\ \bibinfo {author} {\bibfnamefont {R.~G.}\ \bibnamefont
  {Larson}},\ }\bibfield  {title} {\enquote {\bibinfo {title} {Modelling
  hydrodynamic interaction in {Brownian} dynamics: {Simulation} of extensional
  and shear flows of dilute solutions of high molecular weight polystyrene},}\
  }\href@noop {} {\bibfield  {journal} {\bibinfo  {journal} {J. Rheol.}\
  }\textbf {\bibinfo {volume} {48}},\ \bibinfo {pages} {995} (\bibinfo {year}
  {2004})}\BibitemShut {NoStop}%
\bibitem [{\citenamefont {Liu}, \citenamefont {Ashok},\ and\ \citenamefont
  {Muthukumar}(2004)}]{liu:04}%
  \BibitemOpen
  \bibfield  {author} {\bibinfo {author} {\bibfnamefont {S.}~\bibnamefont
  {Liu}}, \bibinfo {author} {\bibfnamefont {B.}~\bibnamefont {Ashok}}, \ and\
  \bibinfo {author} {\bibfnamefont {M.}~\bibnamefont {Muthukumar}},\ }\bibfield
   {title} {\enquote {\bibinfo {title} {Brownian dynamics simulations of
  beadrod-chain in simple shear flow and elongational flow},}\ }\href@noop {}
  {\bibfield  {journal} {\bibinfo  {journal} {Polymer}\ }\textbf {\bibinfo
  {volume} {45}},\ \bibinfo {pages} {1383} (\bibinfo {year}
  {2004})}\BibitemShut {NoStop}%
\bibitem [{\citenamefont {Celani}, \citenamefont {Puliafito},\ and\
  \citenamefont {Turitsyn}(2005)}]{cela:05}%
  \BibitemOpen
  \bibfield  {author} {\bibinfo {author} {\bibfnamefont {A.}~\bibnamefont
  {Celani}}, \bibinfo {author} {\bibfnamefont {A.}~\bibnamefont {Puliafito}}, \
  and\ \bibinfo {author} {\bibfnamefont {K.}~\bibnamefont {Turitsyn}},\
  }\bibfield  {title} {\enquote {\bibinfo {title} {Polymers in linear shear
  flow: A numerical study},}\ }\href@noop {} {\bibfield  {journal} {\bibinfo
  {journal} {Europhys. Lett.}\ }\textbf {\bibinfo {volume} {70}},\ \bibinfo
  {pages} {464} (\bibinfo {year} {2005})}\BibitemShut {NoStop}%
\bibitem [{\citenamefont {Ryder}\ and\ \citenamefont
  {Yeomans}(2006)}]{ryde:06}%
  \BibitemOpen
  \bibfield  {author} {\bibinfo {author} {\bibfnamefont {J.~F.}\ \bibnamefont
  {Ryder}}\ and\ \bibinfo {author} {\bibfnamefont {J.~M.}\ \bibnamefont
  {Yeomans}},\ }\bibfield  {title} {\enquote {\bibinfo {title} {Shear thinning
  in dilute polymer solutions},}\ }\href@noop {} {\bibfield  {journal}
  {\bibinfo  {journal} {J. Chem. Phys.}\ }\textbf {\bibinfo {volume} {125}},\
  \bibinfo {pages} {194906} (\bibinfo {year} {2006})}\BibitemShut {NoStop}%
\bibitem [{\citenamefont {Ripoll}, \citenamefont {Winkler},\ and\ \citenamefont
  {Gompper}(2006)}]{ripo:06}%
  \BibitemOpen
  \bibfield  {author} {\bibinfo {author} {\bibfnamefont {M.}~\bibnamefont
  {Ripoll}}, \bibinfo {author} {\bibfnamefont {R.~G.}\ \bibnamefont {Winkler}},
  \ and\ \bibinfo {author} {\bibfnamefont {G.}~\bibnamefont {Gompper}},\
  }\bibfield  {title} {\enquote {\bibinfo {title} {Star polymers in shear
  flow},}\ }\href@noop {} {\bibfield  {journal} {\bibinfo  {journal} {Phys.
  Rev. Lett.}\ }\textbf {\bibinfo {volume} {96}},\ \bibinfo {pages} {188302}
  (\bibinfo {year} {2006})}\BibitemShut {NoStop}%
\bibitem [{\citenamefont {Sendner}\ and\ \citenamefont {Netz}(2008)}]{send:08}%
  \BibitemOpen
  \bibfield  {author} {\bibinfo {author} {\bibfnamefont {C.}~\bibnamefont
  {Sendner}}\ and\ \bibinfo {author} {\bibfnamefont {R.~R.}\ \bibnamefont
  {Netz}},\ }\href@noop {} {\bibfield  {journal} {\bibinfo  {journal} {EPL}\
  }\textbf {\bibinfo {volume} {81}},\ \bibinfo {pages} {54006} (\bibinfo {year}
  {2008})}\BibitemShut {NoStop}%
\bibitem [{\citenamefont {He}\ \emph {et~al.}(2009)\citenamefont {He},
  \citenamefont {Messina}, \citenamefont {L{\"o}wen}, \citenamefont {Kiriy},
  \citenamefont {Bocharova},\ and\ \citenamefont {Stamm}}]{he:09}%
  \BibitemOpen
  \bibfield  {author} {\bibinfo {author} {\bibfnamefont {G.-L.}\ \bibnamefont
  {He}}, \bibinfo {author} {\bibfnamefont {R.}~\bibnamefont {Messina}},
  \bibinfo {author} {\bibfnamefont {H.}~\bibnamefont {L{\"o}wen}}, \bibinfo
  {author} {\bibfnamefont {A.}~\bibnamefont {Kiriy}}, \bibinfo {author}
  {\bibfnamefont {V.}~\bibnamefont {Bocharova}}, \ and\ \bibinfo {author}
  {\bibfnamefont {M.}~\bibnamefont {Stamm}},\ }\bibfield  {title} {\enquote
  {\bibinfo {title} {Shear-induced stretching of adsorbed polymer chains},}\
  }\href@noop {} {\bibfield  {journal} {\bibinfo  {journal} {Soft Matter}\
  }\textbf {\bibinfo {volume} {5}},\ \bibinfo {pages} {3014} (\bibinfo {year}
  {2009})}\BibitemShut {NoStop}%
\bibitem [{\citenamefont {Zhang}\ \emph {et~al.}(2009)\citenamefont {Zhang},
  \citenamefont {Donev}, \citenamefont {Weisgraber}, \citenamefont {Alder},
  \citenamefont {Graham},\ and\ \citenamefont {de~Pablo}}]{zhan:09}%
  \BibitemOpen
  \bibfield  {author} {\bibinfo {author} {\bibfnamefont {Y.}~\bibnamefont
  {Zhang}}, \bibinfo {author} {\bibfnamefont {A.}~\bibnamefont {Donev}},
  \bibinfo {author} {\bibfnamefont {T.}~\bibnamefont {Weisgraber}}, \bibinfo
  {author} {\bibfnamefont {B.~J.}\ \bibnamefont {Alder}}, \bibinfo {author}
  {\bibfnamefont {M.~G.}\ \bibnamefont {Graham}}, \ and\ \bibinfo {author}
  {\bibfnamefont {J.~J.}\ \bibnamefont {de~Pablo}},\ }\bibfield  {title}
  {\enquote {\bibinfo {title} {Tethered dna dynamics in shear flow},}\
  }\href@noop {} {\bibfield  {journal} {\bibinfo  {journal} {J. Chem. Phys.}\
  }\textbf {\bibinfo {volume} {130}},\ \bibinfo {pages} {234902} (\bibinfo
  {year} {2009})}\BibitemShut {NoStop}%
\bibitem [{\citenamefont {Huang}\ \emph {et~al.}(2010)\citenamefont {Huang},
  \citenamefont {Winkler}, \citenamefont {Sutmann},\ and\ \citenamefont
  {Gompper}}]{huan:10}%
  \BibitemOpen
  \bibfield  {author} {\bibinfo {author} {\bibfnamefont {C.-C.}\ \bibnamefont
  {Huang}}, \bibinfo {author} {\bibfnamefont {R.~G.}\ \bibnamefont {Winkler}},
  \bibinfo {author} {\bibfnamefont {G.}~\bibnamefont {Sutmann}}, \ and\
  \bibinfo {author} {\bibfnamefont {G.}~\bibnamefont {Gompper}},\ }\bibfield
  {title} {\enquote {\bibinfo {title} {Semidilute polymer solutions at
  equilibrium and under shear flow},}\ }\href@noop {} {\bibfield  {journal}
  {\bibinfo  {journal} {Macromolecules}\ }\textbf {\bibinfo {volume} {43}},\
  \bibinfo {pages} {10107} (\bibinfo {year} {2010})}\BibitemShut {NoStop}%
\bibitem [{\citenamefont {Huang}\ \emph {et~al.}(2011)\citenamefont {Huang},
  \citenamefont {Sutmann}, \citenamefont {Gompper},\ and\ \citenamefont
  {Winkler}}]{huan:11}%
  \BibitemOpen
  \bibfield  {author} {\bibinfo {author} {\bibfnamefont {C.-C.}\ \bibnamefont
  {Huang}}, \bibinfo {author} {\bibfnamefont {G.}~\bibnamefont {Sutmann}},
  \bibinfo {author} {\bibfnamefont {G.}~\bibnamefont {Gompper}}, \ and\
  \bibinfo {author} {\bibfnamefont {R.~G.}\ \bibnamefont {Winkler}},\
  }\bibfield  {title} {\enquote {\bibinfo {title} {Tumbling of polymers in
  semidilute solution under shear flow},}\ }\href@noop {} {\bibfield  {journal}
  {\bibinfo  {journal} {EPL}\ }\textbf {\bibinfo {volume} {93}},\ \bibinfo
  {pages} {54004} (\bibinfo {year} {2011})}\BibitemShut {NoStop}%
\bibitem [{\citenamefont {Huang}, \citenamefont {Gompper},\ and\ \citenamefont
  {Winkler}(2012)}]{huan:12}%
  \BibitemOpen
  \bibfield  {author} {\bibinfo {author} {\bibfnamefont {C.-C.}\ \bibnamefont
  {Huang}}, \bibinfo {author} {\bibfnamefont {G.}~\bibnamefont {Gompper}}, \
  and\ \bibinfo {author} {\bibfnamefont {R.~G.}\ \bibnamefont {Winkler}},\
  }\bibfield  {title} {\enquote {\bibinfo {title} {Non-equilibrium relaxation
  and tumbling times of polymers in semidilute solution},}\ }\href@noop {}
  {\bibfield  {journal} {\bibinfo  {journal} {J. Phys.: Condens. Matter}\
  }\textbf {\bibinfo {volume} {24}},\ \bibinfo {pages} {284131} (\bibinfo
  {year} {2012})}\BibitemShut {NoStop}%
\bibitem [{\citenamefont {Lang}, \citenamefont {Obermayer},\ and\ \citenamefont
  {Frey}(2014)}]{lang:14}%
  \BibitemOpen
  \bibfield  {author} {\bibinfo {author} {\bibfnamefont {P.~S.}\ \bibnamefont
  {Lang}}, \bibinfo {author} {\bibfnamefont {B.}~\bibnamefont {Obermayer}}, \
  and\ \bibinfo {author} {\bibfnamefont {E.}~\bibnamefont {Frey}},\ }\href@noop
  {} {\bibfield  {journal} {\bibinfo  {journal} {Phys. Rev. E}\ }\textbf
  {\bibinfo {volume} {89}},\ \bibinfo {pages} {022606} (\bibinfo {year}
  {2014})}\BibitemShut {NoStop}%
\bibitem [{\citenamefont {Nikoubashman}\ and\ \citenamefont
  {Howard}(2017)}]{niko:17}%
  \BibitemOpen
  \bibfield  {author} {\bibinfo {author} {\bibfnamefont {A.}~\bibnamefont
  {Nikoubashman}}\ and\ \bibinfo {author} {\bibfnamefont {M.~P.}\ \bibnamefont
  {Howard}},\ }\href@noop {} {\bibfield  {journal} {\bibinfo  {journal}
  {Macromolecules}\ }\textbf {\bibinfo {volume} {50}},\ \bibinfo {pages} {8279}
  (\bibinfo {year} {2017})}\BibitemShut {NoStop}%
\bibitem [{\citenamefont {Kong}\ \emph {et~al.}(2019)\citenamefont {Kong},
  \citenamefont {Han}, \citenamefont {Chen}, \citenamefont {Cui},\ and\
  \citenamefont {Li}}]{kong:19}%
  \BibitemOpen
  \bibfield  {author} {\bibinfo {author} {\bibfnamefont {X.}~\bibnamefont
  {Kong}}, \bibinfo {author} {\bibfnamefont {Y.}~\bibnamefont {Han}}, \bibinfo
  {author} {\bibfnamefont {W.}~\bibnamefont {Chen}}, \bibinfo {author}
  {\bibfnamefont {F.}~\bibnamefont {Cui}}, \ and\ \bibinfo {author}
  {\bibfnamefont {Y.}~\bibnamefont {Li}},\ }\href@noop {} {\bibfield  {journal}
  {\bibinfo  {journal} {Soft Matter}\ }\textbf {\bibinfo {volume} {15}},\
  \bibinfo {pages} {6353} (\bibinfo {year} {2019})}\BibitemShut {NoStop}%
\bibitem [{\citenamefont {Romo-Uribe}(2021)}]{romo:21}%
  \BibitemOpen
  \bibfield  {author} {\bibinfo {author} {\bibfnamefont {A.}~\bibnamefont
  {Romo-Uribe}},\ }\href@noop {} {\bibfield  {journal} {\bibinfo  {journal} {J.
  Appl. Polym. Sci.}\ }\textbf {\bibinfo {volume} {138}},\ \bibinfo {pages}
  {49712} (\bibinfo {year} {2021})}\BibitemShut {NoStop}%
\bibitem [{\citenamefont {Shee}\ \emph {et~al.}(2021)\citenamefont {Shee},
  \citenamefont {Gupta}, \citenamefont {Chaudhuri},\ and\ \citenamefont
  {Chaudhuri}}]{shee:21}%
  \BibitemOpen
  \bibfield  {author} {\bibinfo {author} {\bibfnamefont {A.}~\bibnamefont
  {Shee}}, \bibinfo {author} {\bibfnamefont {N.}~\bibnamefont {Gupta}},
  \bibinfo {author} {\bibfnamefont {A.}~\bibnamefont {Chaudhuri}}, \ and\
  \bibinfo {author} {\bibfnamefont {D.}~\bibnamefont {Chaudhuri}},\ }\href@noop
  {} {\bibfield  {journal} {\bibinfo  {journal} {Soft Matter}\ }\textbf
  {\bibinfo {volume} {17}},\ \bibinfo {pages} {2120} (\bibinfo {year}
  {2021})}\BibitemShut {NoStop}%
\bibitem [{\citenamefont {Nikoubashman}(2021)}]{niko:21}%
  \BibitemOpen
  \bibfield  {author} {\bibinfo {author} {\bibfnamefont {A.}~\bibnamefont
  {Nikoubashman}},\ }\href@noop {} {\bibfield  {journal} {\bibinfo  {journal}
  {J. Chem. Phys.}\ }\textbf {\bibinfo {volume} {154}},\ \bibinfo {pages}
  {090901} (\bibinfo {year} {2021})}\BibitemShut {NoStop}%
\bibitem [{\citenamefont {Shaqfeh}(2005)}]{shaq:05}%
  \BibitemOpen
  \bibfield  {author} {\bibinfo {author} {\bibfnamefont {E.~S.~G.}\
  \bibnamefont {Shaqfeh}},\ }\href@noop {} {\bibfield  {journal} {\bibinfo
  {journal} {J. Non-Newtonian Fluid Mech.}\ }\textbf {\bibinfo {volume}
  {130}},\ \bibinfo {pages} {1} (\bibinfo {year} {2005})}\BibitemShut {NoStop}%
\bibitem [{\citenamefont {Smith}, \citenamefont {Finzi},\ and\ \citenamefont
  {Bustamante}(1992)}]{smit:92}%
  \BibitemOpen
  \bibfield  {author} {\bibinfo {author} {\bibfnamefont {S.~B.}\ \bibnamefont
  {Smith}}, \bibinfo {author} {\bibfnamefont {L.}~\bibnamefont {Finzi}}, \ and\
  \bibinfo {author} {\bibfnamefont {C.}~\bibnamefont {Bustamante}},\
  }\href@noop {} {\bibfield  {journal} {\bibinfo  {journal} {Science}\ }\textbf
  {\bibinfo {volume} {258}},\ \bibinfo {pages} {1122} (\bibinfo {year}
  {1992})}\BibitemShut {NoStop}%
\bibitem [{\citenamefont {Perkins}\ \emph {et~al.}(1994)\citenamefont
  {Perkins}, \citenamefont {Quake}, \citenamefont {Smith},\ and\ \citenamefont
  {Chu}}]{perk:94}%
  \BibitemOpen
  \bibfield  {author} {\bibinfo {author} {\bibfnamefont {T.~T.}\ \bibnamefont
  {Perkins}}, \bibinfo {author} {\bibfnamefont {S.~R.}\ \bibnamefont {Quake}},
  \bibinfo {author} {\bibfnamefont {D.~E.}\ \bibnamefont {Smith}}, \ and\
  \bibinfo {author} {\bibfnamefont {S.}~\bibnamefont {Chu}},\ }\href@noop {}
  {\bibfield  {journal} {\bibinfo  {journal} {Science}\ }\textbf {\bibinfo
  {volume} {264}},\ \bibinfo {pages} {822} (\bibinfo {year}
  {1994})}\BibitemShut {NoStop}%
\bibitem [{\citenamefont {Doyle}, \citenamefont {Ladoux},\ and\ \citenamefont
  {Viovy}(2000)}]{doyl:00}%
  \BibitemOpen
  \bibfield  {author} {\bibinfo {author} {\bibfnamefont {P.~S.}\ \bibnamefont
  {Doyle}}, \bibinfo {author} {\bibfnamefont {B.}~\bibnamefont {Ladoux}}, \
  and\ \bibinfo {author} {\bibfnamefont {J.-L.}\ \bibnamefont {Viovy}},\
  }\bibfield  {title} {\enquote {\bibinfo {title} {Dynamics of a tethered
  polymer in shear flow},}\ }\href@noop {} {\bibfield  {journal} {\bibinfo
  {journal} {Phys. Rev. Lett.}\ }\textbf {\bibinfo {volume} {84}},\ \bibinfo
  {pages} {4769} (\bibinfo {year} {2000})}\BibitemShut {NoStop}%
\bibitem [{\citenamefont {Lueth}\ and\ \citenamefont
  {Shaqfeh}(2009)}]{luet:09}%
  \BibitemOpen
  \bibfield  {author} {\bibinfo {author} {\bibfnamefont {C.~A.}\ \bibnamefont
  {Lueth}}\ and\ \bibinfo {author} {\bibfnamefont {E.~S.~G.}\ \bibnamefont
  {Shaqfeh}},\ }\href@noop {} {\bibfield  {journal} {\bibinfo  {journal}
  {Macromolecules}\ }\textbf {\bibinfo {volume} {42}},\ \bibinfo {pages} {9170}
  (\bibinfo {year} {2009})}\BibitemShut {NoStop}%
\bibitem [{\citenamefont {Delgado-Buscalioni}(2006)}]{delg:06}%
  \BibitemOpen
  \bibfield  {author} {\bibinfo {author} {\bibfnamefont {R.}~\bibnamefont
  {Delgado-Buscalioni}},\ }\bibfield  {title} {\enquote {\bibinfo {title}
  {Cyclic motion of a grafted polymer under shear flow},}\ }\href@noop {}
  {\bibfield  {journal} {\bibinfo  {journal} {Phys. Rev. Lett.}\ }\textbf
  {\bibinfo {volume} {96}},\ \bibinfo {pages} {088303} (\bibinfo {year}
  {2006})}\BibitemShut {NoStop}%
\bibitem [{\citenamefont {Maier}, \citenamefont {Seifert},\ and\ \citenamefont
  {R{\"a}dler}(2002)}]{maie:02}%
  \BibitemOpen
  \bibfield  {author} {\bibinfo {author} {\bibfnamefont {B.}~\bibnamefont
  {Maier}}, \bibinfo {author} {\bibfnamefont {U.}~\bibnamefont {Seifert}}, \
  and\ \bibinfo {author} {\bibfnamefont {J.~O.}\ \bibnamefont {R{\"a}dler}},\
  }\href@noop {} {\bibfield  {journal} {\bibinfo  {journal} {Europhys. Lett.}\
  }\textbf {\bibinfo {volume} {60}},\ \bibinfo {pages} {622} (\bibinfo {year}
  {2002})}\BibitemShut {NoStop}%
\bibitem [{\citenamefont {Cherstvy}\ and\ \citenamefont
  {Petrov}(2014)}]{cher:14}%
  \BibitemOpen
  \bibfield  {author} {\bibinfo {author} {\bibfnamefont {A.~G.}\ \bibnamefont
  {Cherstvy}}\ and\ \bibinfo {author} {\bibfnamefont {E.~P.}\ \bibnamefont
  {Petrov}},\ }\bibfield  {title} {\enquote {\bibinfo {title} {Modeling dna
  condensation on freestanding cationic lipid membranes},}\ }\href {\doibase
  10.1039/C3CP53433B} {\bibfield  {journal} {\bibinfo  {journal} {Phys. Chem.
  Chem. Phys.}\ }\textbf {\bibinfo {volume} {16}},\ \bibinfo {pages} {2020}
  (\bibinfo {year} {2014})}\BibitemShut {NoStop}%
\bibitem [{\citenamefont {Herold}, \citenamefont {Schwille},\ and\
  \citenamefont {Petrov}(2010)}]{hero:10}%
  \BibitemOpen
  \bibfield  {author} {\bibinfo {author} {\bibfnamefont {C.}~\bibnamefont
  {Herold}}, \bibinfo {author} {\bibfnamefont {P.}~\bibnamefont {Schwille}}, \
  and\ \bibinfo {author} {\bibfnamefont {E.~P.}\ \bibnamefont {Petrov}},\
  }\bibfield  {title} {\enquote {\bibinfo {title} {Dna condensation at
  freestanding cationic lipid bilayers},}\ }\href {\doibase
  10.1103/PhysRevLett.104.148102} {\bibfield  {journal} {\bibinfo  {journal}
  {Phys. Rev. Lett.}\ }\textbf {\bibinfo {volume} {104}},\ \bibinfo {pages}
  {148102} (\bibinfo {year} {2010})}\BibitemShut {NoStop}%
\bibitem [{\citenamefont {Chattopadhyay}\ and\ \citenamefont
  {Marenduzzo}(2007)}]{chat:07}%
  \BibitemOpen
  \bibfield  {author} {\bibinfo {author} {\bibfnamefont {A.~K.}\ \bibnamefont
  {Chattopadhyay}}\ and\ \bibinfo {author} {\bibfnamefont {D.}~\bibnamefont
  {Marenduzzo}},\ }\href@noop {} {\bibfield  {journal} {\bibinfo  {journal}
  {Phys. Rev. Lett.}\ }\textbf {\bibinfo {volume} {98}},\ \bibinfo {pages}
  {088101} (\bibinfo {year} {2007})}\BibitemShut {NoStop}%
\bibitem [{\citenamefont {Lamura}\ and\ \citenamefont
  {Winkler}(2012)}]{lamu:12}%
  \BibitemOpen
  \bibfield  {author} {\bibinfo {author} {\bibfnamefont {A.}~\bibnamefont
  {Lamura}}\ and\ \bibinfo {author} {\bibfnamefont {R.~G.}\ \bibnamefont
  {Winkler}},\ }\bibfield  {title} {\enquote {\bibinfo {title} {Semiflexible
  polymers under external fields confined to two dimensions},}\ }\href@noop {}
  {\bibfield  {journal} {\bibinfo  {journal} {J. Chem. Phys.}\ }\textbf
  {\bibinfo {volume} {137}},\ \bibinfo {pages} {244909} (\bibinfo {year}
  {2012})}\BibitemShut {NoStop}%
\bibitem [{\citenamefont {Hyun}\ \emph {et~al.}(2011)\citenamefont {Hyun},
  \citenamefont {Wilhelm}, \citenamefont {Klein}, \citenamefont {Cho},
  \citenamefont {Nam}, \citenamefont {Ahn}, \citenamefont {Lee}, \citenamefont
  {Ewoldt},\ and\ \citenamefont {McKinley}}]{hyun:11}%
  \BibitemOpen
  \bibfield  {author} {\bibinfo {author} {\bibfnamefont {K.}~\bibnamefont
  {Hyun}}, \bibinfo {author} {\bibfnamefont {M.}~\bibnamefont {Wilhelm}},
  \bibinfo {author} {\bibfnamefont {C.~O.}\ \bibnamefont {Klein}}, \bibinfo
  {author} {\bibfnamefont {K.~S.}\ \bibnamefont {Cho}}, \bibinfo {author}
  {\bibfnamefont {J.~G.}\ \bibnamefont {Nam}}, \bibinfo {author} {\bibfnamefont
  {K.~H.}\ \bibnamefont {Ahn}}, \bibinfo {author} {\bibfnamefont {S.~J.}\
  \bibnamefont {Lee}}, \bibinfo {author} {\bibfnamefont {R.~H.}\ \bibnamefont
  {Ewoldt}}, \ and\ \bibinfo {author} {\bibfnamefont {G.~H.}\ \bibnamefont
  {McKinley}},\ }\bibfield  {title} {\enquote {\bibinfo {title} {A review of
  nonlinear oscillatory shear tests: Analysis and application of large
  amplitude oscillatory shear ({LAOS})},}\ }\href {\doibase
  https://doi.org/10.1016/j.progpolymsci.2011.02.002} {\bibfield  {journal}
  {\bibinfo  {journal} {Prog. Poly. Sci.}\ }\textbf {\bibinfo {volume} {36}},\
  \bibinfo {pages} {1697} (\bibinfo {year} {2011})}\BibitemShut {NoStop}%
\bibitem [{\citenamefont {Rogers}(2018)}]{roge:18}%
  \BibitemOpen
  \bibfield  {author} {\bibinfo {author} {\bibfnamefont {S.}~\bibnamefont
  {Rogers}},\ }\bibfield  {title} {\enquote {\bibinfo {title} {Large amplitude
  oscillatory shear: Simple to describe, hard to interpret},}\ }\href {\doibase
  10.1063/PT.3.3971} {\bibfield  {journal} {\bibinfo  {journal} {Physics
  Today}\ }\textbf {\bibinfo {volume} {71}},\ \bibinfo {pages} {34} (\bibinfo
  {year} {2018})}\BibitemShut {NoStop}%
\bibitem [{\citenamefont {Chen}\ \emph {et~al.}(2005)\citenamefont {Chen},
  \citenamefont {Graham}, \citenamefont {de~Pablo}, \citenamefont {Jo},\ and\
  \citenamefont {Schwartz}}]{chen:05}%
  \BibitemOpen
  \bibfield  {author} {\bibinfo {author} {\bibfnamefont {Y.-L.}\ \bibnamefont
  {Chen}}, \bibinfo {author} {\bibfnamefont {M.~D.}\ \bibnamefont {Graham}},
  \bibinfo {author} {\bibfnamefont {J.~J.}\ \bibnamefont {de~Pablo}}, \bibinfo
  {author} {\bibfnamefont {K.}~\bibnamefont {Jo}}, \ and\ \bibinfo {author}
  {\bibfnamefont {D.~C.}\ \bibnamefont {Schwartz}},\ }\href@noop {} {\bibfield
  {journal} {\bibinfo  {journal} {Macromolecules}\ }\textbf {\bibinfo {volume}
  {38}},\ \bibinfo {pages} {6680} (\bibinfo {year} {2005})}\BibitemShut
  {NoStop}%
\bibitem [{\citenamefont {Thomas}, \citenamefont {DePuit},\ and\ \citenamefont
  {Khomami}(2009)}]{thom:09}%
  \BibitemOpen
  \bibfield  {author} {\bibinfo {author} {\bibfnamefont {D.~G.}\ \bibnamefont
  {Thomas}}, \bibinfo {author} {\bibfnamefont {R.~J.}\ \bibnamefont {DePuit}},
  \ and\ \bibinfo {author} {\bibfnamefont {B.}~\bibnamefont {Khomami}},\
  }\href@noop {} {\bibfield  {journal} {\bibinfo  {journal} {J. Rheol.}\
  }\textbf {\bibinfo {volume} {53}},\ \bibinfo {pages} {275} (\bibinfo {year}
  {2009})}\BibitemShut {NoStop}%
\bibitem [{\citenamefont {Lamura}\ and\ \citenamefont
  {Winkler}(2019)}]{lamu:19}%
  \BibitemOpen
  \bibfield  {author} {\bibinfo {author} {\bibfnamefont {A.}~\bibnamefont
  {Lamura}}\ and\ \bibinfo {author} {\bibfnamefont {R.~G.}\ \bibnamefont
  {Winkler}},\ }\bibfield  {title} {\enquote {\bibinfo {title} {Tethered
  semiflexible polymer under large amplitude oscillatory shear},}\ }\href@noop
  {} {\bibfield  {journal} {\bibinfo  {journal} {Polymers}\ }\textbf {\bibinfo
  {volume} {11}},\ \bibinfo {pages} {737} (\bibinfo {year} {2019})}\BibitemShut
  {NoStop}%
\bibitem [{\citenamefont {Zhou}\ and\ \citenamefont
  {Schroeder}(2016)}]{zhou:16}%
  \BibitemOpen
  \bibfield  {author} {\bibinfo {author} {\bibfnamefont {Y.}~\bibnamefont
  {Zhou}}\ and\ \bibinfo {author} {\bibfnamefont {C.~M.}\ \bibnamefont
  {Schroeder}},\ }\href@noop {} {\bibfield  {journal} {\bibinfo  {journal}
  {Phys. Rev. Fluids}\ }\textbf {\bibinfo {volume} {1}},\ \bibinfo {pages}
  {053301} (\bibinfo {year} {2016})}\BibitemShut {NoStop}%
\bibitem [{\citenamefont {Ripoll}, \citenamefont {Winkler},\ and\ \citenamefont
  {Gompper}(2007)}]{ripo:07}%
  \BibitemOpen
  \bibfield  {author} {\bibinfo {author} {\bibfnamefont {M.}~\bibnamefont
  {Ripoll}}, \bibinfo {author} {\bibfnamefont {R.~G.}\ \bibnamefont {Winkler}},
  \ and\ \bibinfo {author} {\bibfnamefont {G.}~\bibnamefont {Gompper}},\
  }\bibfield  {title} {\enquote {\bibinfo {title} {Hydrodynamic screening of
  star polymers in shear flow},}\ }\href@noop {} {\bibfield  {journal}
  {\bibinfo  {journal} {Eur. Phys. J. E}\ }\textbf {\bibinfo {volume} {23}},\
  \bibinfo {pages} {349} (\bibinfo {year} {2007})}\BibitemShut {NoStop}%
\bibitem [{\citenamefont {Gompper}\ \emph {et~al.}(2009)\citenamefont
  {Gompper}, \citenamefont {Ihle}, \citenamefont {Kroll},\ and\ \citenamefont
  {Winkler}}]{gomp:09}%
  \BibitemOpen
  \bibfield  {author} {\bibinfo {author} {\bibfnamefont {G.}~\bibnamefont
  {Gompper}}, \bibinfo {author} {\bibfnamefont {T.}~\bibnamefont {Ihle}},
  \bibinfo {author} {\bibfnamefont {D.~M.}\ \bibnamefont {Kroll}}, \ and\
  \bibinfo {author} {\bibfnamefont {R.~G.}\ \bibnamefont {Winkler}},\
  }\bibfield  {title} {\enquote {\bibinfo {title} {Multi-particle collision
  dynamics: A particle-based mesoscale simulation approach to the hydrodynamics
  of complex fluids},}\ }\href@noop {} {\bibfield  {journal} {\bibinfo
  {journal} {Adv. Polym. Sci.}\ }\textbf {\bibinfo {volume} {221}},\ \bibinfo
  {pages} {1} (\bibinfo {year} {2009})}\BibitemShut {NoStop}%
\bibitem [{\citenamefont {Kapral}(2008)}]{kapr:08}%
  \BibitemOpen
  \bibfield  {author} {\bibinfo {author} {\bibfnamefont {R.}~\bibnamefont
  {Kapral}},\ }\bibfield  {title} {\enquote {\bibinfo {title} {Multiparticle
  collision dynamics: Simulations of complex systems on mesoscale},}\
  }\href@noop {} {\bibfield  {journal} {\bibinfo  {journal} {Adv. Chem. Phys.}\
  }\textbf {\bibinfo {volume} {140}},\ \bibinfo {pages} {89} (\bibinfo {year}
  {2008})}\BibitemShut {NoStop}%
\bibitem [{\citenamefont {Swope}\ \emph {et~al.}(1982)\citenamefont {Swope},
  \citenamefont {Andersen}, \citenamefont {Berens},\ and\ \citenamefont
  {Wilson}}]{swop:82}%
  \BibitemOpen
  \bibfield  {author} {\bibinfo {author} {\bibfnamefont {W.~C.}\ \bibnamefont
  {Swope}}, \bibinfo {author} {\bibfnamefont {H.~C.}\ \bibnamefont {Andersen}},
  \bibinfo {author} {\bibfnamefont {P.~H.}\ \bibnamefont {Berens}}, \ and\
  \bibinfo {author} {\bibfnamefont {K.~R.}\ \bibnamefont {Wilson}},\ }\bibfield
   {title} {\enquote {\bibinfo {title} {A computer simulation method for the
  calculation of equilibrium constants for the formation of physical clusters
  of molecules: Application to small water clusters},}\ }\href@noop {}
  {\bibfield  {journal} {\bibinfo  {journal} {J. Chem. Phys.}\ }\textbf
  {\bibinfo {volume} {76}},\ \bibinfo {pages} {637} (\bibinfo {year}
  {1982})}\BibitemShut {NoStop}%
\bibitem [{\citenamefont {Allen}\ and\ \citenamefont
  {Tildesley}(1987)}]{alle:87}%
  \BibitemOpen
  \bibfield  {author} {\bibinfo {author} {\bibfnamefont {M.~P.}\ \bibnamefont
  {Allen}}\ and\ \bibinfo {author} {\bibfnamefont {D.~J.}\ \bibnamefont
  {Tildesley}},\ }\href@noop {} {\emph {\bibinfo {title} {Computer Simulation
  of Liquids}}}\ (\bibinfo  {publisher} {Clarendon Press},\ \bibinfo {address}
  {Oxford},\ \bibinfo {year} {1987})\BibitemShut {NoStop}%
\bibitem [{\citenamefont {Malevanets}\ and\ \citenamefont
  {Yeomans}(2000)}]{male:00_1}%
  \BibitemOpen
  \bibfield  {author} {\bibinfo {author} {\bibfnamefont {A.}~\bibnamefont
  {Malevanets}}\ and\ \bibinfo {author} {\bibfnamefont {J.~M.}\ \bibnamefont
  {Yeomans}},\ }\bibfield  {title} {\enquote {\bibinfo {title} {Dynamics of
  short polymer chains in solution},}\ }\href@noop {} {\bibfield  {journal}
  {\bibinfo  {journal} {Europhys. Lett.}\ }\textbf {\bibinfo {volume} {52}},\
  \bibinfo {pages} {231--237} (\bibinfo {year} {2000})}\BibitemShut {NoStop}%
\bibitem [{\citenamefont {Ihle}\ and\ \citenamefont {Kroll}(2001)}]{ihle:01}%
  \BibitemOpen
  \bibfield  {author} {\bibinfo {author} {\bibfnamefont {T.}~\bibnamefont
  {Ihle}}\ and\ \bibinfo {author} {\bibfnamefont {D.~M.}\ \bibnamefont
  {Kroll}},\ }\bibfield  {title} {\enquote {\bibinfo {title} {Stochastic
  rotation dynamics: A {Galilean-invariant} mesoscopic model for fluid flow},}\
  }\href@noop {} {\bibfield  {journal} {\bibinfo  {journal} {Phys. Rev. E}\
  }\textbf {\bibinfo {volume} {63}},\ \bibinfo {pages} {020201(R)} (\bibinfo
  {year} {2001})}\BibitemShut {NoStop}%
\bibitem [{\citenamefont {Lamura}\ \emph {et~al.}(2001)\citenamefont {Lamura},
  \citenamefont {Gompper}, \citenamefont {Ihle},\ and\ \citenamefont
  {Kroll}}]{lamu:01}%
  \BibitemOpen
  \bibfield  {author} {\bibinfo {author} {\bibfnamefont {A.}~\bibnamefont
  {Lamura}}, \bibinfo {author} {\bibfnamefont {G.}~\bibnamefont {Gompper}},
  \bibinfo {author} {\bibfnamefont {T.}~\bibnamefont {Ihle}}, \ and\ \bibinfo
  {author} {\bibfnamefont {D.~M.}\ \bibnamefont {Kroll}},\ }\bibfield  {title}
  {\enquote {\bibinfo {title} {Multiparticle collision dynamics: Flow around a
  circular and a square cylinder},}\ }\href@noop {} {\bibfield  {journal}
  {\bibinfo  {journal} {Europhys. Lett.}\ }\textbf {\bibinfo {volume} {56}},\
  \bibinfo {pages} {319--325} (\bibinfo {year} {2001})}\BibitemShut {NoStop}%
\bibitem [{\citenamefont {Ladoux}\ and\ \citenamefont {Doyle}(2000)}]{lado:00}%
  \BibitemOpen
  \bibfield  {author} {\bibinfo {author} {\bibfnamefont {B.}~\bibnamefont
  {Ladoux}}\ and\ \bibinfo {author} {\bibfnamefont {P.~S.}\ \bibnamefont
  {Doyle}},\ }\bibfield  {title} {\enquote {\bibinfo {title} {Stretching
  tethered dna chains in shear flow},}\ }\href@noop {} {\bibfield  {journal}
  {\bibinfo  {journal} {Europhys. Lett.}\ }\textbf {\bibinfo {volume} {52}},\
  \bibinfo {pages} {511} (\bibinfo {year} {2000})}\BibitemShut {NoStop}%
\bibitem [{\citenamefont {Harnau}, \citenamefont {Winkler},\ and\ \citenamefont
  {Reineker}(1995)}]{harn:95}%
  \BibitemOpen
  \bibfield  {author} {\bibinfo {author} {\bibfnamefont {L.}~\bibnamefont
  {Harnau}}, \bibinfo {author} {\bibfnamefont {R.~G.}\ \bibnamefont {Winkler}},
  \ and\ \bibinfo {author} {\bibfnamefont {P.}~\bibnamefont {Reineker}},\
  }\href@noop {} {\bibfield  {journal} {\bibinfo  {journal} {J. Chem. Phys.}\
  }\textbf {\bibinfo {volume} {102}},\ \bibinfo {pages} {7750} (\bibinfo {year}
  {1995})}\BibitemShut {NoStop}%
\bibitem [{\citenamefont {Arag\'{o}n}\ and\ \citenamefont
  {Pecora}(1985)}]{arag:85}%
  \BibitemOpen
  \bibfield  {author} {\bibinfo {author} {\bibfnamefont {S.~R.}\ \bibnamefont
  {Arag\'{o}n}}\ and\ \bibinfo {author} {\bibfnamefont {R.}~\bibnamefont
  {Pecora}},\ }\href@noop {} {\bibfield  {journal} {\bibinfo  {journal}
  {Macromolecules}\ }\textbf {\bibinfo {volume} {18}},\ \bibinfo {pages} {1868}
  (\bibinfo {year} {1985})}\BibitemShut {NoStop}%
\bibitem [{\citenamefont {Rouse}(1953)}]{rous:53}%
  \BibitemOpen
  \bibfield  {author} {\bibinfo {author} {\bibfnamefont {P.~E.}\ \bibnamefont
  {Rouse}},\ }\href@noop {} {\bibfield  {journal} {\bibinfo  {journal} {J.
  Chem. Phys.}\ }\textbf {\bibinfo {volume} {21}},\ \bibinfo {pages} {1272}
  (\bibinfo {year} {1953})}\BibitemShut {NoStop}%
\bibitem [{\citenamefont {Doi}\ and\ \citenamefont {Edwards}(1986)}]{doi:86}%
  \BibitemOpen
  \bibfield  {author} {\bibinfo {author} {\bibfnamefont {M.}~\bibnamefont
  {Doi}}\ and\ \bibinfo {author} {\bibfnamefont {S.~F.}\ \bibnamefont
  {Edwards}},\ }\href@noop {} {\emph {\bibinfo {title} {The Theory of Polymer
  Dynamics}}}\ (\bibinfo  {publisher} {Clarendon Press},\ \bibinfo {address}
  {Oxford},\ \bibinfo {year} {1986})\BibitemShut {NoStop}%
\end{thebibliography}

%

\end{document}